\documentclass[onecolumn]{elsart3p}
\usepackage{amssymb}
\usepackage{amsfonts}
\usepackage{amsmath}
\usepackage{graphicx}
\usepackage{subfigure}
\usepackage{epsfig}
\usepackage{color}

\input{tcilatex}

\begin{document}

\begin{frontmatter}%

\title{Analytic continuation of Taylor series and the boundary
value problems of some nonlinear ordinary differential equations}%
\author{S.\ Abbasbandy$^a$}, \author{C.\ Bervillier$^b$}\ead{claude.bervillier@lmpt.univ-tours.fr}%

%
\address{$^a$ Department of Mathematics, Science and Research Branch, Islamic Azad University, Tehran, Iran\\}

\address{$^b$ Laboratoire de Math\'{e}matiques et Physique Th\'{e}orique,\\ UMR 6083 (CNRS),\\
F\'ed\'eration Denis Poisson,\\
Universit\'{e} Fran\c{c}ois Rabelais,\\
Parc de Grandmont, 37200 Tours, France}%

\begin{abstract}
We compare and discuss the\ respective efficiency of three methods 
(with two variants for each of them), based
respectively on Taylor (Maclaurin)
series, Pad\'{e} approximants and conformal mappings, for solving
quasi-analytically a two-point boundary value problem of a nonlinear
ordinary differential equation (ODE).
Six configurations of ODE and boundary conditions are successively
considered according to the increasing difficulties that they present. 
After having indicated that the Taylor series method 
almost always requires the recourse to analytical continuation procedures to be efficient, we use
the complementarity of the two remaining methods (Pad\'{e} and conformal mapping) to illustrate their respective
advantages and limitations. We emphasize the importance of the existence of
solutions with movable singularities for the efficiency of the methods,
particularly for the so-called Pad\'{e}-Hankel method. (We show that this
latter method is equivalent to pushing a movable pole to infinity.) For each
configuration, we determine the singularity distribution (in the complex
plane of the independent variable)\ of the solution sought and show how
this distribution controls the efficiency of the two methods. In general the
method based on Pad\'{e} approximants is easy to use and robust but may be awkward in some circumstances whereas the conformal mapping method is a very fine method which should
be used when high accuracy is required.%
\end{abstract}%


\end{frontmatter}%

\section{Introduction}

The use of expansions about the initial boundary (Taylor series) for solving
an initial value (Cauchy) problem of an ordinary differential equation (ODE)
is a well known procedure. Eventually an analytic continuation is needed to
enlarge the limited range of validity of the series. It is less known that
Taylor series may be used also to solve two-point boundary value 
problems (BVP) for ODEs. An old example is the Blasius problem \cite{6552} (a non
linear ODE with specific conditions at the two boundaries of the domain $%
\mathcal{D}^{+}=[0,+\infty )$ of the independent variable, see section \ref%
{Blasius}). In this case, the Taylor series is a Maclaurin series
and, contrary to a Cauchy problem, one value is lacking at the origin (the\
unknown connection parameter\footnote{%
We limit ourselves to problems where only one initial value is lacking (the
connection parameter $k$) which is the value at the origin of the function, or of one of its derivative. It is to be determined in order
to satisfy one condition at the second boundary located at infinity.} $k$)
to make the series explicit. The\ Blasius method consists in first
expressing the generic solution as a power series in the independent
variable about the origin, the coefficients, $c_{n}\left( k\right) $, of which,
depend on the connection parameter. In a second step, the missing value $%
k^{\ast }$ of $k$ is determined by adjusting its value so that the sum of
the series (or one of its derivative) matches the condition at the second
boundary located at infinity\footnote{%
In fact Blasius made use of a supplementary asymptotic expansion of the
solution at the infinity point and matched the two expansions in a middle
region of $\mathcal{D}^{+}$. This raises the supplementary question of the
convergence of the asymptotic expansion.}. Of course, as in the Cauchy
problem, the Blasius series has a finite radius of convergence and, at best,
\ the \textquotedblleft \textsl{method can yield results of limited accuracy}%
\textquotedblright\ (Weyl \cite{7099}).

To improve the Blasius procedure, one must call for some analytic
continuation of the series. But, unless one uses a computer and a
symbolic calculation software, this is practically impossible because $k$ is
unknown. Moreover, with a nonlinear ODE, the possible appearance of
spontaneous and/or movable singularities (the locations of which may depend
on $k$) reduces, in an a priori unverifiable way, the magnitude of the
radius of convergence of the Maclaurin series. This is why it is relatively
recently that advances in the computing power have rendered practical the
efficient use of several quasi-analytical methods for solving BVPs 
\cite{6188,6194,6335,6201,6110,6398,6394} (though the basic ideas are not
always new).

On general grounds, in a well posed BVP, one is seeking the
unique solution that has some good analyticity properties, at least in 
$\mathcal{D}^{+}$. The methods based on Taylor series for solving a BVP
are implicit in that one attempts to determine the only value $%
k^{\ast }$ (the determination of an explicit solution in terms of the
independent variable being left to a subsidiary step which often is
obvious). For each method, there are two main variants for determining $k^{\ast }$,
according to whether the condition at the second boundary is explicitly
imposed or not. We shall name them respectively ``explicit'' and ``minimal''.

\begin{description}
\item[``Explicit'' procedure:] In the first case, one constructs an
explicit candidate solution (depending generically on $k)$ which is valid in
the whole domain of definition $\mathcal{D}^{+}$ of the solution sought.
The value $k^{\ast }$ is\ then determined by imposing explicitly the
condition at the second boundary (auxiliary condition). The explicit approximate solution is then
obtained by substituting $k^{\ast }$ for $k$ in the candidate solution. This
procedure always works in principle (provided the candidate solution is well
chosen) but in practice it may become very cumbersome due to the necessity of
working with complete expressions.

\item[``Minimal'' procedure:] The second case is simpler since it amounts to imposing that the solution is merely analytically  compatible with 
a particular class of functions having some (hopefully appropriate) analyticity properties in $\mathcal{D}^{+}$,
such as generalized hypergeometric
functions \cite{6188} or rational functions --the Pad\'{e}-Hankel method 
\cite{6194,6335,6201}-- or which satisfy a first-order polynomial ODE \cite%
{6110} or even the Taylor series of the generic solution itself --the simplistic method  \cite{6499}. As in the ``explicit'' procedure, one first determines the specific function (hypergeometric or rational or etc... ) 
by the usual conventional matching with the coefficients of the Taylor series 
of the generic solution truncated at a given order $M$. Then, assuming that the function so constructed is analytically compatible with the solution of interest, one pushes the matching rules to the next order $M+1$ all other things being equal.
The auxiliary condition so obtained is used to
determine $k^*$. With such a procedure the
requirement that the solution does satisfy the condition
at the second boundary is not explicitly imposed.  The ``minimal'' procedure is based on the hope that the solution sought
has properties of analyticity similar to the function (or representation) considered. It
 may be
 efficient only when this solution is an isolated
particular solution of the ODE\footnote{%
It may also belong to an infinite discrete set of solutions defined in $%
\mathcal{D}^{+}$.} defined in $\mathcal{D}^{+}$ (e.g. the envelope of a
family of solutions having a movable singularity\footnote{%
We explicitly show in section \ref{Pade} that the Pad\'{e}-Hankel method is
equivalent to forcing the localization at infinity of a movable singularity
(when it exists).}). In that case, imposing only specific
conditions of analyticity in the domain $\mathcal{D}^{+}$ may be sufficient
to distinguish that solution among the others. Such occasions are not always
realized, as, for example, with the Blasius problem which cannot be solved
with ``minimal'' procedures, but ``explicit'' procedures work, see section \ref%
{Blasius} (this explains why, using the Pad\'{e}-Hankel method --the ``minimal'' variant of the Padé method-- Amore and Fern\'andez have failed to determine the solution of that problem \cite{6190}).
\end{description}

In this paper, after a brief presentation of the rough Taylor series method (and its ``minimal'' variant named the simplistic method Bervillier \cite{6499}),
we discuss and compare the efficiency of two complementary
methods of using Taylor series for solving a BVP. They
rely upon the use of Pad\'{e} approximants and of
conformal mappings respectively.
Of course, each of these methods has a ``minimal'' variant, named the Pad\'{e}-Hankel method (Fern\'andez and coworkers \cite{6194,6335,6201}) in the first case and with no particular name in the second case \cite{6188,6499}.
Our choice is founded by their apparent easiness and efficiency relatively
to other methods but also on their complementarity. As ``explicit'' procedures,
the two methods are not new since they have already been proposed and used
long ago to solve a BVP (e.g., see van Dyke \cite[pp. 205--213]{7112} under
the respective name of rational fractions and Euler transforms). Only the
smallness of the computing power of that time had limited their
applications. More recent  (``explicit'') applications of the methods for solving a BVP
may also be found in Boyd \cite{7089,7094}. The use of ``minimal'' variants is less common.

Our aim is twofold. Firstly we want to illustrate some conditions of
application of the two methods with respect to the wide variety of domains
of analyticity that one may encounter with a nonlinear ODE. The underlying
idea is to try and propose some rules which may help in a priori
determining whether the methods have a chance to succeed or not. Secondly we
want to actually show that the two methods are complementary. In fact, the
Pad\'{e} method, when it works, is efficient, provided relatively low orders
are concerned, when the desired accuracy requires the consideration of high
orders of the series then the more sophisticated mapping method takes over.
However, in turn, Pad\'{e} approximants may be used to determine with a
sufficient accuracy the two adjustable parameters involved in the mapping
method.

A short review on the use of Taylor series for solving a BVP for
an example of a linear ODE (the eigenvalue problem of the anharmonic
oscillator) may be found in \cite{6499}. Here we consider explicitly
several nonlinear ODEs chosen for the different analyticity properties of
their solutions in order to show (a small but significant part of) the
variety of complexities that one may encounter. It is a commonplace to say
that the disposition of the singularities in the complex plane controls the
convergence of a Taylor series, it is less common to explicitly show this
phenomenon in actual examples. The ODEs that we consider are the following
(by order of complexity, each of them is accompanied by particular initial
conditions not mentioned in this enumeration):

\begin{itemize}
\item The Polchinski fixed point equation \cite{354} in the local potential
approximation (an exact, or nonperturbative, renormalization group equation,
for introductory reviews see e.g. \cite{4595}). This is an easy example of
second order ODEs having a single (most probably unique) non trivial
solution defined in $\mathcal{D}^{+}$ (it is the envelope of the general
solution --a two-parameter family-- having a movable singularity).
``Minimal'' procedures (not the simplistic method however) succeed in determining this particular solution (due
to its uniqueness). Although analytic in $\mathcal{D}^{+}$, the (particular)
solution of interest has singularities in the complex plane of the
independent variable located exclusively on the negative real axis and the
radius of convergence of its Maclaurin series is not small. The analytic
continuations presently considered are then very efficient (see section \ref%
{Polchinski}).

\item The Thomas-Fermi equation \cite{7103,7104} for the neutral atom, a
second order ODE (and boundary conditions) with facilities similar to the
previous case (general solution with a movable singularity, the solution
of the BVP is the envelope of a family of such singular solutions). The main
difference is that the singularities of the solution sought are no
longer confined to the negative real axis but are arranged in the half-plane
of the negative real axis (lhs half-plane). The convergence of the summation
is then made more difficult than in the preceding case, but the procedures
considered (of any kind) work well again (see section \ref{Thomas}).

\item The ODE which controls the flow of a third grade fluid in a porous
half space \cite{6638,6634}. With this equation the solution of interest has
a Maclaurin series with a smaller radius of convergence and the general
solution has an essential movable singularity at infinity (the leading part
of the movable singularity is no longer a pole located at a movable point)
whereas the solution sought (most probably unique) goes to zero.
Procedures of any kind still work but they are not very efficient compared
to the previous cases, it is preferable to consider the problem from the
second boundary using a domain compactification: $\mathcal{D}^{+}=[0,+\infty
)\rightarrow \left[ 1,0\right] $ via an essential change of definition of
the independent variable (Ahmad \cite{6634}), for some values of the parameters of
the ODE, the new Maclaurin series then converges for $k=k^{\ast }$ (see
section \ref{ThirdGrade}).

\item The Falkner-Skan flow equation. For the values of the parameters of
this ODE that we consider, the singularities of the solution sought are
disposed along an arc of a circle around the origin (open toward the positive
axis) with singularities that get in the half plane of the positive real
axis (rhs half-plane) of the independent variable. This solution
appears as the envelope of a two-parameter (particular) solution of the ODE with a
movable singularity, it is (most probably) the unique solution defined in $%
\mathcal{D}^{+}$ which satisfies the initial conditions. Procedures of any
kind work, despite the slow convergence of the resummations (see section \ref%
{FS}).

\item The Blasius problem. This is a particular case of the preceding ODE.
The previous two-parameter (particular) solution with a movable singularity
disappears to leave room for a continuum of solutions defined in $\mathcal{D}%
^{+}$. The ``minimal'' procedures fail to determine a solution. Fortunately,
the combination of a scaling property with the particular symmetry of the
problem allows to transform the BVP into a simple Cauchy problem
and makes the singularities of the solution sought disposed in the half
plane of the negative real axis of the effective independent variable,
consequently ``explicit'' procedures may be used with actual efficiency (see
section \ref{Blasius}).

\item The convex solution of the Blasius equation with more general initial
conditions. It is the same kind of configuration as previously but the
problem cannot be reduced to a Cauchy problem and the singularities of the
solution of interest get in the right-hand-side half plane so that it is
difficult to sum the Maclaurin series efficiently. Even the ``explicit''
procedures have difficulty to give an approximate solution (see section \ref%
{Convex}).
\end{itemize}

The organisation of the paper is as follows. We first shortly present the
Taylor series method and the two analytic continuations that we consider. We
then discuss each example of ODE independently. We finally conclude. In an appendix, we explicitly illustrate how to systematically implement a local expansion of a (general or particular) solution of an ODE.

\section{Taylor series method and analytic continuation}

\subsection{Taylor series method}

This is an old and well known method for solving formally an ODE and several
detailed presentations may be found (see for example \cite{wiki1}%
). We limit ourselves here to a brief introduction of its application to the
BVP.

Suppose that the function $g\left( z\right) $ satisfies a second order ODE:%
\begin{equation}
F(g^{\prime \prime },g^{\prime },g,z)=0\,,  \label{eq:EDO1}
\end{equation}%
in which a symbol $^{\prime }$ denotes a derivative with respect to the
independent variable $z$.

Because the general solution depends on two arbitrary constants, the BVP
is to find (if it exists) the solution in a range\footnote{%
In fact, we shall have $z\in \mathcal{D}^{+}=[0,+\infty )$ with the examples
explicitly considered below.} $z\in \left[ a,b\right] $ which satisfies,
e.g. the two following boundary conditions\footnote{%
The conditions could be, as well, expressed on the derivatives of $g$ or
mixed, provided that their number be equal to the order of the ODE.}:%
\begin{equation}
\left. 
\begin{array}{c}
g\left( a\right) =A\,, \\ 
g\left( b\right) =B\,,%
\end{array}%
\right\}  \label{eq:set2points}
\end{equation}%
in which $A$ and $B$ are two given constants. It would be easier to
determine the solution corresponding to two conditions given at only one
boundary (the Cauchy problem), say, for example:%
\begin{equation}
\left. 
\begin{array}{c}
g\left( a\right) =A\,, \\ 
g^{\prime }\left( a\right) =k\,.%
\end{array}%
\right\}  \label{eq:Cauchy}
\end{equation}%
In fact, the problem (\ref{eq:set2points}) amounts to solve the Cauchy
problem (\ref{eq:Cauchy}) for an unknown (missing) value $k^{\ast }$ of $k$
such that the condition $g\left( b\right) =B$ be satisfied. If a unique
solution exists, one may solve this problem numerically using, e.g., a
shooting method and a Newton-Raphson procedure of adjustment. But presently
one is rather interested in quasi-analytical methods based on the
representation of the solution via Taylor series.

To this end, one first Taylor expands the solution about the boundary $a$
(Maclaurin series if $a=0$):%
\begin{equation*}
g\left( z\right) =\sum_{n=0}^{\infty }c_{n}\left( z-a\right) ^{n}\,, 
\end{equation*}
and determines the coefficients $c_{n}$ in terms of $A$ and $k$, such that
the equation (\ref{eq:EDO1}) be satisfied order by order in powers of $%
\left( z-a\right) $. One gets that way the set $c_{n}\left( A,k\right) $
(with $c_{0}=A$, $c_{1}=k$, etc...). It remains to sum the resulting series
in order to satisfy the condition at the second boundary which determines
the value $k^{\ast }$ of $k$:%
\begin{equation}
\sum_{n=0}^{\infty }c_{n}\left( A,k^{\ast }\right) \left( b-a\right)
^{n}=B\,.  \label{eq:ConditionFinale}
\end{equation}%
Finally the global solution of the problem is (formally) given by:%
\begin{equation*}
g^{\ast }\left( z\right) =\sum_{n=0}^{\infty }c_{n}\left( A,k^{\ast }\right)
\left( z-a\right) ^{n}\,. 
\end{equation*}

Of course the Taylor series method remains formal as long as one does not
know whether the series converges or not. If it does, the method will only
provide an approximate result since only series of a\ limited number of
terms are considered in practice:%
\begin{equation}
g_{M}\left( z\right) =\sum_{n=0}^{M}c_{n}\left( A,k\right) \left( z-a\right)
^{n}\,.  \label{eq:TaylorM}
\end{equation}

At this stage we distinguish the ``explicit'' and the ``minimal''  variants of the Taylor method.
\subsubsection{``Explicit'' procedure (Taylor method)}
The condition at the second boundary is explicitly imposed, namely:
\begin{equation}
\sum_{n=0}^{M}c_{n}\left( A,k\right) \left( b-a\right) ^{n}=B\,, \label{eq:TaylorExpli}
\end{equation}%
which provides the auxiliary condition to determine $k^*$. This method is effectively used in section \ref{ThirdGrade}
where a useful change of function and a compactification of the initial BVP makes the resulting
Taylor series convergent. But,
in general,
one encounters a difficulty when the second boundary goes to infinity ($b\rightarrow+\infty$) because the lhs of (\ref{eq:TaylorExpli}
) goes to
$\pm\infty$ according to the sign of the last term. It is then better to switch to the ``minimal'' variant.
\subsubsection{``Minimal'' procedure (simplistic method)}
The auxiliary condition is obtained by imposing that the next term of the truncated Taylor series vanishes:
\begin{equation}
c_{M+1}\left( A,k\right)=0\,. \label{eq:MOP}
\end{equation}

 That is equivalent to assuming that, for $k=k^*$ (ideally determined when $M\rightarrow\infty$),
the series converges within the boundaries and the hope is that the condition (\ref{eq:MOP}) will be sufficient to determine approximately $k^*$ for relatively small values of $M$. In the case where $b=+\infty$ this would mean
that the series of the desired solution has an infinite radius of convergence. In practice
this is a so strong assumption that it is almost never true: the radius of convergence is most often limited by the presence of singularities in the complex plane
of the independent variable $z$. Nevertheless it occurs sometimes that the simplistic method provides reasonable approximated estimations of $k^*$ (Margaritis et al \cite{3478}). For that two conditions are required:

\begin{enumerate}
\item the solution sought is the unique analytic solution (or belongs to a finite set of separate analytic solutions). 
This is a condition required with any ``minimal'' procedure.
\item the other solutions have movable singularities which must be (temporarily) located within the domain of convergence of the Taylor series whereas $k$
is already close to $k^*$\footnote{Actually, $k^*$ may be seen as the value of $k$ for which the movable singularities are pushed to infinity, if those singularities
are already outside the domain of convergence of the Taylor series when $k$ is not close to $k^*$, the simplistic method cannot work.}.
\end{enumerate}
Though 
the simplistic method does not work in general, it is so simple that it should be systematically tried before considering any other method.
Unfortunately, it does not work with the set of ODEs considered below.

When the Taylor series does not converge (almost always), one must call for
analytic continuation methods such as Pad\'{e} approximants or conformal mappings%
\footnote{%
There are other possible analytic continuations. For example, the simplest
one would be to consider the continuous
analytic continuation method which consists in performing successive Taylor
expansions each within its supposed circle of convergence until the second
boundary be reached. But this method is not well adapted when the second boundary
is located at infinity.} (see section \ref{Accel}).

\subsection{General remarks}
With the explicit examples of ODE and the methods considered in the present article, the
auxiliary equations
 for $k$ [similar to (\ref{eq:TaylorExpli}) or (\ref{eq:MOP})]
that we consider below are polynomial equations. The number of candidate values for $k^*$ (the zeros of the polynomial) equals the degree of the polynomial which grows with $M$. Even after having eliminating the complex zeros at a given order $M$, there is no possibility to distinguish the right value $k^*$ among all the remaining real zeros. But if the solution of the BVP is unique, one may expect to observe, when $M$ grows, a clear convergence of one (and only one\footnote{%
This is still true if there is a discrete set of isolated solutions which
satisfy the conditions, then one should observe several distinct convergent series of (real) zeros, e.g.,
see the eigenvalue problem of the anharmonic oscillator \cite{6499} where
the generic eigenvalue parameter plays the role of the connection parameter.}) series
 of (real) zeros toward the right value $k^{\ast }$.
 When the method works, the convergent series may be distinguished aleady when $M$ is small so that it is easy to locate and follow it for larger values of $M$  (without having to determine all the other --uninteresting-- zeros). Figure 4 of ref \cite{6499} illustrates how such convergent series of zeros may be easily located and followed. When no such series of zeros appears for small values of $M$, then the method works hardly or does not work at all. It is precisely one of the aims of this paper to illustrate how and why such situations may arise.
 
The
high degree of the polynomials to be treated makes it necessary to
determine $k^{\ast }$ numerically\footnote{%
But with an unlimited accuracy if one uses a symbolic calculation software.}
so that\ the final expression of the solution is not completely put under an
analytical form, the approximate method is called quasi-analytic.

\subsection{Analytic continuation methods \label{Accel}}

\subsubsection{Rational functions (Pad\'{e} approximants)\label{Pade}}

\paragraph{``Explicit'' procedure (Pad\'{e} method)}

Given the Taylor series (\ref{eq:TaylorM}) of finite order $M$, one tries to
approximate its sum by a rational function:%
\begin{equation}
g_{M}\left( z\right) =\frac{P\left( z-a\right) }{Q\left( z-a\right) }\,,
\label{eq:RationalFunction}
\end{equation}%
in which $P\left( z\right) $ and $Q(z)$ are polynomial functions of
respective degree $p$ and $q$. The coefficients of the monomials are
unambiguously determined as functions of $A$ and $k$ from (\ref{eq:TaylorM})
so that:%
\begin{eqnarray}
P\left( z-a\right) &=&Q\left( z-a\right) \,g_{M}\left( z\right) +O\left[
(z-a)^{M+1}\right] \,,  \label{eq:Pad1} \\
Q\left( 0\right) &=&1\,,  \label{eq:Pad2} \\
p+q &=&M\,.  \label{eq:Pad3}
\end{eqnarray}

The representation (\ref{eq:RationalFunction}) of $g_{M}\left( z\right) $ is
also called the Pad\'{e} approximant $[p,q]$ of $g_{M}\left( z\right) $. It
is a kind of analytic continuation of the truncated Taylor series (e.g., see Baker and/or Baker and Graves-Morris
\cite{7110}).

The solution of the boundary problem (\ref{eq:EDO1}, \ref{eq:set2points}) is
then hopefully obtained by determining the values of $k$ for which the
condition at the second boundary is satisfied:%
\begin{equation}
\frac{P\left( b-a\right) }{Q\left( b-a\right) }=B\,,  \label{eq:Pade1kind}
\end{equation}%
which is again a polynomial equation for $k$.

The use of the explicit condition (\ref{eq:Pade1kind}) at the second
boundary to determine $k^{\ast }$ makes the procedure ``explicit''. In principle $%
g^{\ast }\left( z\right) $ is unique and one should observe only one convergent series of 
zeros of the polynomial in $k$ as $M$ grows. Notice that, for a given $M$,
the pair $\left( p,q\right) $ is not unique and may be chosen according to
the expected behavior\ of $g^{\ast }\left( z\right) $ at the second
boundary. In practice one may try several pairs $\left( p,q\right) $ to
obtain the best convergence of the series of zeros toward $k^{\ast }.$

The ``explicit'' procedure requires the explicit calculation (in terms of
the unknown $k)$ of all the monomials of the rational function (\ref%
{eq:RationalFunction}) and this is generally very time consuming. Moreover,
it may occur that the behavior of the solution at the second boundary is not
compatible with a rational function and one must utilize some artifact to
impose an efficient condition [see, e.g. eq. (\ref{eq:BlasiusPAD})]. In addition, a Pad\'e approximant may sometimes be ill defined \cite{7110} so that it cannot be an explicit representation of the solution and the method accidentally fails.

The recourse to the
``minimal'' procedure allows to bypass these difficulties at least in the process of determining an approximate value of $k^*$.

\paragraph{``Minimal'' procedure (Pad\'{e}-Hankel method)}

According to the presentation given in the introduction, one imposes that
the polynomials $P$ and $Q$ as
defined at order $M$ using (\ref{eq:Pad1}--\ref{eq:Pad3}) match again eq. (\ref{eq:Pad1})
at next order (with  $M\rightarrow M+1$).
The linear system of equations which then defines the coefficients of $P$
and $Q$ becomes overdetermined and leaves room for a definition of $k$ via
the vanishing of the determinant $\det (\hat{T})$ of a Toeplitz matrix%
\footnote{%
The matrix $\hat{T}_{i,j}$ may be rewritten under the form of a Hankel
matrix $\hat{H}_{i,l}$ by a redefinition of the index $l=q-j$. The
reference to a Toeplitz matrix is more convenient with regards to the
discussion presented in section \ref{PHMovable}.} $\hat{T}_{i,\,l}$\
constructed with the coefficients $c_{n}\left( A,k\right) $ \cite%
{6194,6335,6201}:%
\begin{equation}
\hat{T}_{i,\,j}=c_{q+\omega +1+i-j}\qquad \left( i=0,\cdots ,q;j=0,\cdots
,q\right) \,,  \label{eq:MatrixH}
\end{equation}%
in which $\omega =p-q$.

Finally, the condition%
\begin{equation}
\det (\hat{T})=0\,,  \label{eq:detH}
\end{equation}%
provides a polynomial equation for $k$ (the degree of which grows with $M$).

Since the condition at the second boundary is not imposed explicitly, one is
not assured (in particular if $b$ is finite) that there is a unique solution
and the set of zeros of (\ref{eq:detH}) may or may not contain an
approximate value of $k^{\ast }$ (see section \ref{EDOS}). Let us show
that, if the second boundary is located at infinity ($b=+\infty $), the Pad%
\'{e}-Hankel method has a particular significance.

\paragraph{The Pad\'{e}-Hankel method and the removal of a movable
singularity\label{PHMovable}}

Suppose that $b=+\infty $ and that the generic\footnote{%
Depending on the unknown connection parameter $k$.} solution of the ODE,
corresponding to the initial conditions under study, has a singularity
located at the movable point $z_{0}$, then the solution $g^{\ast }\left(
z\right) $, defined in $\mathcal{D}^{+}$, is the envelope
of that family of (singular at $z_{0}$) solutions. This envelope may be seen
as the limit $z_{0}\rightarrow +\infty $ of the family. Consequently, to
determine $k^{\ast }$, one may proceed as follows.

Instead of imposing that the ratio $P/Q$ should also reproduce the series at
next order $g_{M+1}\left( z\right) $ one requires that the rational
approximation has a singularity (a pole) located at some $z_{0}$, i.e.:%
\begin{equation*}
Q\left( z_{0}\right) =0\,, 
\end{equation*}%
this induces a modification of the matrix (\ref{eq:MatrixH}) which then
reads:%
\begin{eqnarray*}
\tilde{T}_{i,\,j} &=&c_{q+\omega +1+i-j}\qquad \left( i=0,\cdots
,q-1;j=0,\cdots ,q\right) \,, \\
\tilde{T}_{q,\,j} &=&z_{0}^{j}\qquad \left( j=0,\cdots ,q\right) \,.
\end{eqnarray*}

By imposing that $\det (\tilde{T})=0$ one obtains a polynomial equation for $%
k$ which depends on $z_{0}$. When $z_{0}$ becomes very large,
$\det (\tilde{T})$ is dominated by the largest power of $z_{0}$
($z_{0}^q$) so that, when $z_{0}\rightarrow+\infty$, the effective auxiliary condition
asymptotically reduces to the vanishing of a minor of the matrix $\tilde{T}$ which is
 the determinant of the
following reduced matrix:%
\begin{equation*}
\tilde{T}_{i,\,j}=c_{q+\omega +1+i-j}\qquad \left( i=0,\cdots
,q-1;j=0,\cdots ,q-1\right) \,. 
\end{equation*}%

That is nothing but the Toeplitz matrix (\ref{eq:MatrixH}) considered at
the preceding order $(q\rightarrow q-1)$ but with $\omega \rightarrow \omega
+1$.

Hence the Pad\'{e}-Hankel method (for determining $k^{\ast }$) is closely
related to the process of sending a movable pole at infinity. Its efficiency
thus depends on the existence of such a movable singularity.

\subsubsection{Conformal mappings\label{Mapping}}

In the following we assume that the two boundaries are $a=0$ and $b=+\infty $%
. Let us consider a continuation of the independent variable $z$ in the
complex.

The conformal mapping method for solving a BVP has been introduced
by Bervillier et al \cite{6319} to which we refer the reader for more details.

If one knows that the solution (i.e. when $k=k^{\ast }$) is
analytic in the right-hand side interior of an angular sector as drawn in
figure 1 of \cite{6319}, then one can effectuate the following conformal
transformation ($R>0$, $\alpha>0$):

\begin{equation}
z\rightarrow w=\frac{\left( 1+z/R\right) ^{1/\alpha }-1}{\left( 1+z/R\right)
^{1/\alpha }+1}\,,  \label{eq:Map}
\end{equation}%
which maps the region of the $z$-plane formed by the interior of the angular sector the vertex of which is located on the negative real axis at $-R$ and of angle $\alpha \pi /2$ measured in radians
into the interior of the unit circle centered at the origin of the $w$-plane
so that $z=+\infty $ corresponds to $w=+1$ (whereas $z=0$ corresponds to $%
w=0 $).

The transformation (\ref{eq:Map}) is better understood as the result of the  following sequence of tranformations:
\begin{enumerate}
\item 
$z\rightarrow z_{1}=z/R$,

\item
$z_{1}\rightarrow z_{2}=z_{1}+1$, shift of the vertex of the angular sector onto the origin,

\item 
$z_{2}\rightarrow z_{3}=\left( z_{2}\right) ^{1/\alpha }$, mapping of the interior of the angular sector onto the positive half plane,

\item
$z_{3}\rightarrow w=(z_{3}-1)/(z_{3}+1)$, mapping of the positive half plane onto the unit disc $\left\vert w\right\vert<1 $.
\end{enumerate}

This conformal mapping is merely a generalization of the Euler transform which
corresponds (see, e.g. van Dyke \cite[p. 208]{7112}) to set $\alpha
=1 $ and $R=1/2 $ (the rhs half plane with boundary located at $z=-1/2$).

If the interior of the angular sector is a region of analyticity of the
original function $g\left( z\right) $ then the series (\ref{eq:TaylorM})
converges there. Under the analytic continuation this convergence is
conveyed to the whole unit disc for the series: 
\begin{equation}
\tilde{g}_{M}\left( w\right) =\sum_{i=0}^{M}u_{i}\left( A,k\right) \,w^{i}\,,
\label{eq:SerieAfterMap}
\end{equation}%
obtained by expanding, within the original series (\ref{eq:TaylorM}), the
relation inverse of (\ref{eq:Map}):%
\begin{equation}
z=R\left[ \left( \frac{1+w}{1-w}\right) ^{\alpha }-1\right] \,.
\label{eq:MapInverse}
\end{equation}
Before looking at the way one may determine the values of $R$ and $\alpha$, let us define the two variants of the mapping method.
\paragraph{``Explicit'' procedure}

The condition at the second (here infinite) boundary may be imposed by
setting $w=1$ in (\ref{eq:SerieAfterMap}):%
\begin{equation}
\tilde{g}_{M}\left( 1\right) =\sum_{i=0}^{M}u_{i}\left( A,k\right) \,=B\,.
\label{eq:Borne2Mapping}
\end{equation}

\paragraph{``Minimal'' procedure}

The condition (\ref{eq:Borne2Mapping}) indicates that the mapping method is
naturally an ``explicit'' procedure. It may be used also formally by simply
imposing that the last term of the series (\ref{eq:SerieAfterMap}) vanishes:%
\begin{equation}
u_{M}\left( A,k\right) =0\,. \label{eq:formalMap}
\end{equation}

As in the case of the simplistic method (see eq. [\ref{eq:MOP}]), one forces the convergence of the series,
but this time the series of the solution of interest actually converges provided the conformal mapping has been well chosen.
 This gives a simple and
light version of the mapping method for determining $k^{\ast }$ however the
gain compared to the use of the ``explicit'' procedure is not as large as that
one may observe with Pad\'{e} approximants. Moreover, as mentioned previously in
section \ref{Pade}, the ``minimal'' procedure may not work at all (see
section \ref{EDOS}).

\paragraph{Choice of the parameters $R$ and $\protect\alpha $}
The determination of the parameters $\alpha $ and $R$ depends on the
distribution of the singularities of $g^{\ast }\left( z\right) $ in the
complex plane which is a priori unknown. In practice one may try
several values of $\alpha $ and $R$ beginning with sufficiently small values
so as to get a convergent series of zeros for $k$ in (\ref{eq:Borne2Mapping}) or (\ref{eq:formalMap}). Then
this rough estimate of $k^{\ast }$ may be used to determine a better value
of $R$ by the d'Alembert or the Cauchy rule \cite{6188,6499} because, most often\footnote{%
Not always because the maximal value of $R$ may be larger than the radius of
convergence of the series (see section \ref{ConvexAnal}).} the maximal value
of $R$ is the radius of convergence of the series (\ref{eq:TaylorM}) for $%
k=k^{\ast }$. 

Actually the radius of convergence of the Taylor series is fixed by the position in the complex plane of the independent variable of the singularity
the closest to the origin (remind that $a=0$). If one assumes that this position is known to be $z_s=x+i y$, then one has: $R=\sqrt{x^2+y^2}$ and $\alpha=2 \arcsin\left(|y|/h\right)/\pi $ (in which $h=\sqrt{(R+x)^2+y^2}$). 

We shall show that Pad\'{e} approximants may be
used to determine the values of $R$  and $\alpha$ (in a way similar to that of Andersen and Geer \cite{7150}).

\section{Explicit examples of ODEs \label{EDOS}}

In the following we shall\ generically call $g^{\ast }\left( z\right) $ the
solution of the BVP of interest, it corresponds to a (hopefully
unique) value $k=k^{\ast }$ of the connection parameter which will be the
\textquotedblleft missing\textquotedblright\ value of $g$ or of one of its
derivative at the origin. The assumed domain of definition of $g^{\ast
}\left( z\right) $ is $\mathcal{D}^{+}=\left[ 0,+\infty \right) $, but
sometimes solution defined in $\mathcal{D}^{-}=\left[ 0,-\infty \right) $
will be also encountered.

\subsection{The Polchinski fixed point equation in the local potential
approximation \label{Polchinski}}

The problem is to find the potential $V\left( x\right) =V^{\ast }\left(
x\right) $ which satisfies the following second order ODE and boundary
conditions:%
\begin{eqnarray}
V^{\prime \prime }-V^{\prime 2}-\frac{1}{2}xV^{\prime }+3V &=&0\,,
\label{eq:Pol} \\
V^{\prime }\left( 0\right) &=&0\,,  \label{eq:PolCond1} \\
V^{\prime \prime }\left( \infty \right) &=&1\,.  \label{eq:PolCond2}
\end{eqnarray}

The connection parameter associated to that problem is:%
\begin{equation}
V\left( 0\right) =\gamma \,,  \label{eq:PolConn}
\end{equation}%
and one must determine the value $\gamma =\gamma ^{\ast }$ which corresponds
to $V^{\ast }\left( x\right) $.

It is convenient to work with the derivative $V^{\prime }\left( x\right) $
and to perform the following change of variable: 
\begin{eqnarray*}
f\left( x\right) &=&V^{\prime }\left( x\right) \,, \\
f\left( x\right) &=&x\,g\left( x^{2}\right) \,, \\
z &=&x^{2}\,,
\end{eqnarray*}%
then the system (\ref{eq:Pol}--\ref{eq:PolCond2}) is equivalent to:%
\begin{eqnarray}
4z\,g^{\prime \prime }-4z\,g\,g^{\prime }-2g^{2}+\left( 6-z\right)
\,g^{\prime }+2\,g &=&0\,,  \label{eq:Pol2} \\
g\left( +\infty \right) &=&1\,,  \label{eq:PolCond22}
\end{eqnarray}%
with the new connection parameter defined as 
\begin{equation}
k=g\left( 0\right) =-3\gamma \,.  \label{eq:PolCond12}
\end{equation}

Notice that, compared to the original BVP (\ref{eq:Pol}--\ref%
{eq:PolCond2}), there is no explicit initial condition (e.g. on $g^{\prime
}\left( 0\right) $), it is replaced by the following condition of
consistency on (\ref{eq:Pol2}) at $z=0$ [a regular singular point of (\ref%
{eq:Pol2})]:%
\begin{equation*}
g^{\prime }\left( 0\right) =\frac{k\left( k-1\,\right) }{3}\,, 
\end{equation*}%
which is automatically fulfilled when $g\left( z\right) $ is expressed as a
Maclaurin series the first terms of which are:

\begin{equation*}
g\left( z\right) =k+\frac{1}{3}(k-1)k\,z+\frac{1}{60}(k-1)k(8k-1)\,z^{2}+%
\frac{1}{630}(k-1)k^{2}(34k-13)\,z^{3}+O\left( z^{4}\right) \,. 
\end{equation*}

It is relatively easy to verify the following (In appendix A we show how to obtain expansions like (\ref{eq:Polgsing}, \ref{eq:Polasy}).:

\begin{enumerate}
\item eq. (\ref{eq:Pol2}) admits two trivial solutions analytic in $\mathcal{%
D}^{+}$:

\begin{enumerate}
\item $g_{1}^{\ast }\left( z\right) =0$

\item $g_{2}^{\ast }\left( z\right) =1$
\end{enumerate}

\item eq. (\ref{eq:Pol2}) admits, locally to $z_{0}$, the following singular
solution (for a discussion of the reasons which lead to consider such
expansions as solution of an ODE, see, e.g. \cite{7156,7155}):%
\begin{equation}
g_{\mathrm{sing}}\left( z\right) =-\frac{2}{\left( z-z_{0}\right) }+\frac{2-%
\text{$z_{0}$}}{4\text{$z_{0}$}}+\left( z-z_{0}\right) \left\{ C\,+O\left[
\log \left( \left\vert z-z_{0}\right\vert \right) ,\left( z-z_{0}\right) %
\right] \right\} \,,  \label{eq:Polgsing}
\end{equation}%
which depends on the two arbitrary constants $z_{0}$ and $C$ (movable
singularity). Provided the expansion (\ref{eq:Polgsing}) converges, it
represents locally the general solution of (\ref{eq:Pol2}) which, thus, has
necessarily a domain of definition smaller than $\mathcal{D}^{+}$ except,
perhaps, when $z_{0}\rightarrow +\infty $. Consequently, $g^{\ast }\left(
z\right) $ does not belong to that two-parameter family, it is a particular
solution of (\ref{eq:Pol2}).

\item eq. (\ref{eq:Pol2}) admits a particular (one-parameter) solution
which, asymptotically for large $z,$ behaves as:%
\begin{equation}
g\left( z\right) \underset{z\rightarrow \infty }{\simeq }1+G\,z^{-2/5}+\frac{%
G^{2}}{5}z^{-4/5}+O\left( z^{-6/5}\right) \,,  \label{eq:Polasy}
\end{equation}%
in which $G$ is an arbitrary constant.
\end{enumerate}

The asymptotic one-parameter solution (\ref{eq:Polasy}) satisfying (\ref%
{eq:PolCond22}), indicates the possible existence of the solution $g^{\ast
}\left( z\right) $ as the envelope of the two-parameter family of locally
singular solutions (\ref{eq:Polgsing}). By reversing the BVP
(starting from the infinite boundary and with the connection parameter $G)$,
it is likely that a solution associated to any value of $G$ will display a
singularity somewhere before reaching the origin, except for a particular
value $G^{\ast }$ (the envelope) which corresponds to $g^{\ast }\left(
z\right) $.

In conclusion, if we except the two trivial solutions $g_{1}^{\ast }\left(
z\right) $ and $g_{2}^{\ast }\left( z\right) ,$ the solution $g^{\ast
}\left( z\right) $ is the only solution defined in $\mathcal{D}%
^{+}$. In practice it is the envelope ($z_{0}\rightarrow +\infty $) of the
two-parameter family of locally singular solutions (\ref{eq:Polgsing}) with $%
C$ (continuously) adapted to the initial conditions.

Since it is not singular on the positive real axis, one may expect that the
domain of analyticity of $g^{\ast }\left( z\right) $ is sufficiently large
to allow an analytic continuation of its Maclaurin series from the origin up
to infinity. Moreover, since it belongs to a discrete set of nonsingular
solutions [i.e., the set $\left\{ g_{1}^{\ast }\left( z\right) ,g_{2}^{\ast
}\left( z\right) ,g^{\ast }\left( z\right) \right\} $], ``minimal''
procedures should work. This is confirmed a posteriori with the help of Pad%
\'{e} approximants and conformal mappings.

With $50$ terms in the Maclaurin series ($M=50$) $k^{\ast }$ has been
estimated by Amore and Fern\'andez in \cite{6190} using the Pad\'{e}-Hankel method yielding $20$
digits of accuracy\footnote{%
The authors had estimated $k^{\ast }-1$ instead of $k^{\ast }.$}:%
\begin{equation}
k^{\ast }\approx -0.22859820243702192438\,.  \label{eq:Polkstar}
\end{equation}

Since, in a ``minimal'' procedure like the Pad\'{e}-Hankel method, the
condition at infinity is not imposed explicitly, it is almost indifferent
whether one considers a basis of diagonal or a basis of sub diagonal
approximants, in addition the two expected trivial solutions are also
determined by observing (among the numerous values of $k$ proposed as $M$
grows) the presence of two additional convergent series of zeros towards 0 and 1.

According to the considerations of section \ref{PHMovable} the success of
the Pad\'{e}-Hankel method is most certainly related to the possibility of
moving to infinity ($z_{0}\rightarrow \infty $) of the movable singularity (%
\ref{eq:Polgsing}). But the broad accuracy of (\ref{eq:Polkstar}) is also
surely a consequence of some particular analyticity properties of $g^{\ast
}\left( z\right) $. Let us look at them.

Knowing the value (\ref{eq:Polkstar}), even with fewer digits, one may
explicitly use Pad\'{e} approximants (diagonal or sub diagonal) on the
Maclaurin series of $g^{\ast }\left( z\right) $ (with $k\approx k^{\ast }$)
with a view to determine the singularity distribution of $g^{\ast }\left(
z\right) $. The procedure is not new (e.g., see Andersen and Geer \cite{7150}) and may be described as
follows. The denominator of the rational function constructed from the
series has some complex zeros which often reproduce the singular points of
the actual function that it approximates. In general such singular points
appear as the zeros (of the denominator) which remain stable as the order of
the series grows (the spurious zeros being unstable). When the function to
be approximated has a cut or a branch point it is simulated by an
accumulation of poles on the cut. This is precisely what happens in the
present case. For $k\approx -0.228598$, the zeros of the denominator of the
Pad\'{e} approximants of a Maclaurin series (truncated to some varying value
of $M$) accumulate exclusively on the negative side of the real axis of $z$.
The stable zero (under changing the order $M$ of the series) the closest to
the origin controls the radius of convergence and gives $R=5.7217$.
Consequently, the angular sector of apparent analyticity would be the whole
plane cut on the negative real axis and this corresponds to setting $\alpha
=2$ in eq. (\ref{eq:Map}). The values: 
\begin{equation}
R=5.7217,\qquad \alpha =2\,,  \label{eq:PolMaxRal}
\end{equation}%
are thus the theoretically optimal values for the two parameters of the
conformal mapping method. The latter value of $R$ corroborates that obtained
previously by using the d'Alembert rule \cite{6188} on the terms of the
Maclaurin series for $k\approx k^{\ast }$.

In ref. \cite{6319}, the conformal mapping method has provided a better
estimate of $k^{\ast }$ than (\ref{eq:Polkstar}) using values of $R$ and $%
\alpha $ close to the maximal values\footnote{%
When the order of the series is not very high, the maximal values of $R$ and 
$\alpha $ are not necessarily the values which give the best apparent
convergence of the conformal mapping method. One observes, however, a
tendency to favor the maximal values at very high order.\label{PolMax}} (%
\ref{eq:PolMaxRal}) and with $M=120$:%
\begin{equation}
k^{\ast }\approx -0.228598202437021924373656107397047188424054430086.
\label{eq:Polkstar2}
\end{equation}

It is important to realize that, though one does not know a priori the exact
value of $k^{\ast }$, this (high) accuracy is internally controlled by the
method itself. Indeed the convergence of the method may only be spoiled by
ill controlled singularities. If the convergence is manifest, then the
validity of the assumed properties of analyticity is confirmed and in turn
the number of stable digits is secured.

With the present case of the Polchinski ODE (\ref{eq:Pol2}), the Pad\'{e}%
-Hankel method is slightly more efficient than the conformal mapping method
for relatively low orders. However it is very time consuming due to the
necessity of calculating larger and larger determinants as the order $M$
grows. The conformal mapping method is a bit less rapidly convergent but
finally takes over because the calculations may be easily pushed to higher
orders.

Let us mention also that with the conformal mapping method, both
procedures (``explicit'' and ``minimal'', see section \ref{Accel}) yield roughly
the same accuracy. As for the Pad\'{e} method, though the behavior (\ref%
{eq:Polasy}) at the second boundary of the solution cannot be reproduced by
a rational function one can impose an explicit condition. Actually, due to the
condition (\ref{eq:PolCond22}), one can obtain an estimate of $k^{\ast }$ by
using diagonal Pad\'{e} approximants to represent the solution and by merely
imposing that the ratio of the highest degrees of the respective polynomials 
$P$ and $Q$ be equal to 1. But considering the great efficiency of the
``minimal'' procedure, the recourse to the ``explicit'' version of the Pad%
\'{e} method is not pertinent here.

Let us emphasize that once $k^{\ast }$ has been determined within some
accuracy, each of the two methods (except sometimes accidentally with a Pad\'e approximant) provides an explicit analytical
(approximate) expression for the solution: a rational fraction (a Pad\'{e} approximant) or a convergent series in powers of $w\left(
z\right) $ as given by (\ref{eq:Map}). This remark is obviously valid in any
occasion where the methods work.

\subsection{The Thomas-Fermi equation for the neutral atom \label{Thomas}}

The Thomas-Fermi problem for the neutral atom \cite{7103,7104} writes in
terms of a function $u\left( x\right) $:%
\begin{eqnarray*}
u^{\prime \prime }-\sqrt{\frac{u^{3}}{x}} &=&0\,, \\
u\left( 0\right) &=&1\,, \\
u\left( +\infty \right) &=&0\,,
\end{eqnarray*}%
with the connection parameter $k=u^{\prime }\left( 0\right) $.

As in the previous example, it is convenient to perform a change of function
and of variable:%
\begin{eqnarray*}
g\left( z\right) &=&\sqrt{u\left( z^{2}\right) }\,, \\
z &=&x^{1/2}\,,
\end{eqnarray*}%
the\ original problem is then equivalent to:%
\begin{eqnarray}
z\,\left[ g\,g^{\prime \prime }+g^{\prime }{}^{2}\right] -g\,g^{\prime
}-2\,z^{2}\,g^{3} &=&0\,,  \label{eq:TF2} \\
g\left( 0\right) &=&1\,,  \label{eq:TF21} \\
g\left( +\infty \right) &=&0\,,  \label{eq:TF2inf}
\end{eqnarray}%
and the connection parameter is:%
\begin{equation*}
k=g^{\prime \prime }\left( 0\right) \,. 
\end{equation*}

The first terms of the Maclaurin series of $g\left( z\right) $ are:

\begin{equation*}
g\left( z\right) =1+\frac{kz^{2}}{2}+\frac{2z^{3}}{3}-\frac{k^{2}z^{4}}{8}-%
\frac{2kz^{5}}{15}+\frac{1}{18}\left( \frac{9k^{3}}{8}-1\right) z^{6}+\frac{%
6k^{2}z^{7}}{35}+O\left( z^{8}\right) \,. 
\end{equation*}

The Thomas-Fermi equation has much more been mathematically studied than the
previous one. In particular it is known, by a theorem of Mambriani \cite%
{7111} (see \cite{7102}) that $g^{\ast }\left( z\right) $ is unique.
Sommerfeld \cite{7105} has calculated its behavior for large $z$, and one
may easily verify (see appendix A) that (\ref{eq:TF2}) admits the following asymptotic
expansion as a one-parameter solution:%
\begin{equation}
g\left( z\right) \underset{z\rightarrow +\infty }{\simeq }\frac{12}{z^{3}}+%
\frac{G}{z^{s}}+\cdots ;\quad s=\sqrt{73}-4\,, \label{eq:TFasy}
\end{equation}%
in which $G$ is the only arbitrary constant. As in the Polchinski case, it
exists a value $G^{\ast }$ which should correspond to $g^{\ast }\left(
z\right) $ and, presumably, it is the envelope of a family of singular
solutions satisfying the initial condition (\ref{eq:TF21}).

Indeed one may easily show that, similarly to (\ref{eq:Polgsing})\ for the
Polchinski case  (see appendix A), the general solution of (\ref{eq:TF2}) has a two-parameter
movable singularities which, locally to any $z_{0},$ behaves as: 
\begin{eqnarray}
&&g_{\mathrm{sing}}\left( z\right) \underset{z\rightarrow z_{0}}{\simeq }%
\frac{5}{\text{$z_{0}$}}\frac{1}{(z-\text{$z_{0}$})^{2}}-\frac{20}{9\ \text{$%
z_{0}$}^{2}}\frac{1}{(z-\text{$z_{0}$})}+\frac{35}{36\text{$z_{0}$}^{3}}-%
\frac{1015}{2916\text{$z_{0}$}^{4}}(z-\text{$z_{0}$})  \notag \\
&&-\frac{1175}{104976\text{$z_{0}$}^{5}}(z-\text{$z_{0}$})^{2}+\frac{43805}{%
157464\text{$z_{0}$}^{6}}(z-\text{$z_{0}$})^{3}  \notag \\
&&-\frac{136949335}{238085568\text{$z_{0}$}^{7}}(z-\text{$z_{0}$})^{4}+\frac{%
2338092205}{2142770112\text{$z_{0}$}^{8}}(z-\text{$z_{0}$})^{5}  \notag \\
&&-\frac{189476597645}{77139724032\text{$z_{0}$}^{9}}(z-\text{$z_{0}$})^{6}+%
\frac{37136484275}{4132485216\text{$z_{0}$}^{10}}(z-\text{$z_{0}$})^{7} 
\notag \\
&&+(z-\text{$z_{0}$})^{8}\left( C+\frac{113234508800\log (\left\vert \text{$%
z_{0}$}-z\right\vert )}{4261625379z_{0}^{11}}\right) +\cdots \,,
\label{eq:TFgsing}
\end{eqnarray}%
in which $z_{0}$ and $C$ are two arbitrary constants\footnote{%
The late presence of the logarithm in this expansion prevents the ODE from
having the Painlev\'{e} property (for a review see, e.g. \cite{7156,7155}).
Notice also that the explicit calculation of all the terms of (\ref%
{eq:TFgsing}) to reveal the presence of the arbitrary constant $C$ in front
of the monomial $(z-z_{0}$$)^{8},$ is not necessary. Such a term is
sometimes named a resonance. Once the leading term has been determined, it
is sufficient to find the value of the degree $n$ of an arbitrary
additional monomial which linearly contributes to zero in the ODE. In the
present case, one finds $n=8$. (See appendix A.)}. As in the previous case, it represents
locally the general solution of (\ref{eq:TF2}) which, thus, has necessarily
a domain of definition smaller than $\mathcal{D}^{+}$ except, perhaps, when $%
z_{0}\rightarrow \infty $.

Recalling section \ref{PHMovable}, we are not surprised that
the Pad\'{e}-Hankel method provides an easy determination of $k^{\ast }$.
With $M\approx 60$ Amore and Fern\'andez \cite{6190} (see also Fern\'andez \cite{6532,6381}) using this method have
obtained the following estimate: 
\begin{equation}
k^{\ast }\approx -1.58807102261137532\,.  \label{eq:kstarPHTF1}
\end{equation}

In the circumstances and as in the Polchinski case of section \ref%
{Polchinski}, with regards to the efficiency of the ``minimal'' procedure, it
is not necessary to make use of the ``explicit'' procedure with Pad\'{e}\
approximants to determine $k^{\ast }$. But with the knowledge of $k^{\ast }$%
, one may construct a rational fraction (only once) to obtain an explicit
(approximate) solution. As explained in section \ref{Polchinski}, that
approximate explicit solution may be used to roughly determine the
distribution of the singularities of $g^{\ast }\left( z\right) $.

For $k\approx -1.5880710$ and values of $M$ up to $120$, we observe,
compared to the Polchinski case,\ a wider dispersions of the stable
singularities, but they remain confined to $\func{Re}\left( z\right) <0$
with a cut on the real axis starting at $z=-0.6658$. The maximal angular
sector of analyticity appears to be characterized by the value $\alpha =1.1$%
. Hence the values:%
\begin{equation}
R=0.6658,\quad \alpha =1.10\,,  \label{eq:TFMaxRal}
\end{equation}%
are the theoretically maximal values for the two parameters of the conformal
mapping method. This corresponds to a smaller range of analyticity than in
the Polchinski case.

According to the remark of footnote \ref{PolMax}, for practical efficiency,
we have chosen $R=0.65$ and the convenient $\alpha =1$ which are close to
the maximal values (\ref{eq:TFMaxRal}) and with $M=200$ we have obtained the
following estimate of $k^{\ast }$:%
\begin{equation*}
k^{\ast }\approx -1.5880710226113753127189\pm 7\times 10^{-22}\,, 
\end{equation*}%
which is only slightly\footnote{%
Compared to the Polchinski case for which (\ref{eq:Polkstar2}) is much more
accurate than (\ref{eq:Polkstar}) whereas $M=120$.} better than the result (%
\ref{eq:kstarPHTF1}) previously obtained by using the Pad\'{e}-Hankel
method. Actually, at low orders, the calculations clearly show a greater
efficiency of the Pad\'{e}-Hankel method compared to the relatively poor
results of the conformal mapping method (which provides a better estimate
because high orders may be treated).

We observe also a smaller rate of convergence of the two methods, compared
to the Polchinski case. The effective value of $\alpha $ being smaller than
in the this latter case (where the maximal allowed values $\alpha =2$ is
reached), one may conclude that the (relative) greater difficulties
encountered with the Thomas-Fermi ODE, is a direct consequence of a less
favorable distribution of singularities.

\subsection{The third grade fluid in a porous half space\label{ThirdGrade}}

In this section we consider the equation of the flow of a third grade fluid
in a porous half space which has been investigated by Hayat et al \cite{6638} and Ahmad \cite{6634}. The
formulation of this problem in terms of $f\left( z\right) $ is as:

\begin{eqnarray}
f^{\prime \prime }+b_{1}f^{\prime }{}^{2}f^{\prime \prime }-\frac{b_{1}c}{3}ff^{\prime }{}^{2}-cf &=0,&
\label{eq:TGF} \\
\ f(0) &=1,&  \label{eq:TGF1} \\
\ f(+\infty ) &=0.&  \label{eq:TGF2}
\end{eqnarray}%
in which $b_{1}$ and $c$ are two constants.

Using a rescaling of the form:%
\begin{equation*}
f\left( z\right) =\frac{1}{\sqrt{b_{1}c}}\,g\left( \sqrt{c}\,z\right) \,, 
\end{equation*}%
the problem (\ref{eq:TGF}--\ref{eq:TGF2}) may be written as follows:%
\begin{eqnarray}
\,g^{\prime \prime }+g^{\prime 2}g^{\prime \prime }-\frac{1}{3}\,gg^{\prime
2}-\,g &=&0\,,  \label{eq00bis} \\
g\left( 0\right) &=&A=\sqrt{b_{1}c}\,,  \label{eq01bis} \\
g\left( +\infty \right) &=&0\,.  \label{eq02bis}
\end{eqnarray}

The connection parameter is again:%
\begin{equation*}
k=g^{\prime }\left( 0\right) \,. 
\end{equation*}

For information, the particular values $b_{1}=0.6$ and $c=0.5$ considered in 
\cite{6634} had yielded the following estimate (using a shooting method \cite%
{6634}):%
\begin{equation*}
f^{\ast \prime }\left( 0\right) \approx -0.678301\,, 
\end{equation*}%
which, in our conventions, corresponds to 
\begin{eqnarray}
A &=&\sqrt{b_{1}c}\approx 0.5477225575\,,  \label{eqA} \\
k^{\ast } &=&\sqrt{b_{1}}f^{\ast \prime }\left( 0\right) \approx -0.525410\,.
\notag
\end{eqnarray}%
Having fixed the values of $b_{1}$ and $c$ to the values indicated above,
the first terms of the Maclaurin series of $g\left( z\right) $ read as
follows:

\begin{equation*}
g\left( z\right) =\sqrt{\frac{3}{10}}+kz+\frac{\left( k^{2}+3\right) z^{2}}{2%
\sqrt{30}\left( k^{2}+1\right) }+\frac{k\left( k^{2}+3\right) \left(
5k^{4}+10k^{2}+3\right) z^{3}}{90\left( k^{2}+1\right) ^{3}}+O\left(
z^{4}\right) \,. 
\end{equation*}

It is easy to verify that the ODE (\ref{eq00bis}) does not admit,\ locally
to $z_{0},$ any solution of the form\footnote{%
Since the ODE is autonomous, it is sufficient to look for a solution of the
form $\mathcal{A}\,z^{\alpha }$ (because of the translation invariance under
the change $z\rightarrow z+z_{1}$).}:%
\begin{equation}
g_{\mathrm{sing}}\left( z\right) \underset{z\rightarrow z_{0}}{\simeq }%
\mathcal{A}\text{ }\left( z-z_{0}\right) ^{s}\,,  \label{eq:gsingpower}
\end{equation}%
but for $s=1$ which makes $g_{\mathrm{sing}}\left( z\right) $ analytic.

Actually the general solution has a movable essential singularity at
infinity since for large $z$, (\ref{eq00bis}) asymptotically admits the\
following two-parameter family of behavior as solution  (see appendix A):

\begin{equation}
g\left( z\right) \underset{z\rightarrow +\infty }{\simeq }C_{1}e^{\frac{z}{%
\sqrt{3}}}-\left( \frac{\sqrt{3}}{\text{$C_{1}$}}z+C_{2}\right) e^{-\frac{z}{%
\sqrt{3}}}+O(z\,e^{-3\frac{z}{\sqrt{3}}})\,,  \label{eq:AhmadSing}
\end{equation}%
in which $C_{1}$ and $C_{2}$ are two arbitrary constants. Notice that the
limit $C_{1}\rightarrow 0$ in (\ref{eq:AhmadSing}) is singular, this
indicates the possible existence of another kind of solution.

Actually, there is a one-parameter family of solutions of (\ref{eq00bis})
which, for large $z>0$, behaves as:%
\begin{equation}
g\left( z\right) \underset{z\rightarrow +\infty }{\simeq }Ge^{-z}-\frac{G^{3}%
}{12}e^{-3z}\,.  \label{eqserinfini0}
\end{equation}

This kind of solution (which goes to zero at the second boundary) should
correspond to $g^{\ast }\left( z\right) $ for a particular value $G^{\ast }$%
. It is the envelope of the solutions which, for $z\rightarrow +\infty $ go
to $\pm \infty $ according to the sign of $C_{1}$ in (\ref{eq:AhmadSing}).
Indeed it corresponds to the singular limit $C_{1}\rightarrow 0$ of the
general solution corresponding to (\ref{eq:AhmadSing}).

Despite the probable uniqueness\footnote{%
Actually there is also the same kind of solution defined in $\mathcal{D}^{-}$
that corresponds to the value $k=-k^{\ast }$ and its presence is well
observed as a  convergent series of zeros in the Pad\'{e}-Hankel method. This solution
is not observed with the ``minimal'' version of the mapping method since only 
$\mathcal{D}^{+}$ is actually mapped onto the unit circle.} of $g^{\ast
}\left( z\right) $ as a solution of (\ref{eq00bis}) defined in $\mathcal{D}%
^{+}$, the Pad\'{e}-Hankel method works with difficulty yielding the
following poor estimate of $k^{\ast }$:%
\begin{equation}
k^{\ast }\approx -0.525\,,  \label{eq:Ahmadkstarpade}
\end{equation}%
with the order of the Maclaurin series limited (to save time) to $M=17$.
This poor result is in agreement with the absence of movable pole and
confirms the considerations of section \ref{PHMovable} on the nature of the
Pad\'{e}-Hankel method.

The ``explicit'' procedure with diagonal Pad\'{e} approximants has not improved
the result.

One may associate this poor behavior with the fact that essential
singularities located at infinity seem to be analytic when they are seen
from the origin, they are more difficult to detect than poles or cuts
located at a finite distance from the origin.

Despite the poor accuracy of (\ref{eq:Ahmadkstarpade}), Pad\'{e}
approximants may be used to\ roughly give the distribution of singularities
of $g^{\ast }\left( z\right) $ in the complex plane of $z$ (see figure \ref%
{fig9}). 

\begin{figure}[tbp]
\begin{center} 
\includegraphics*{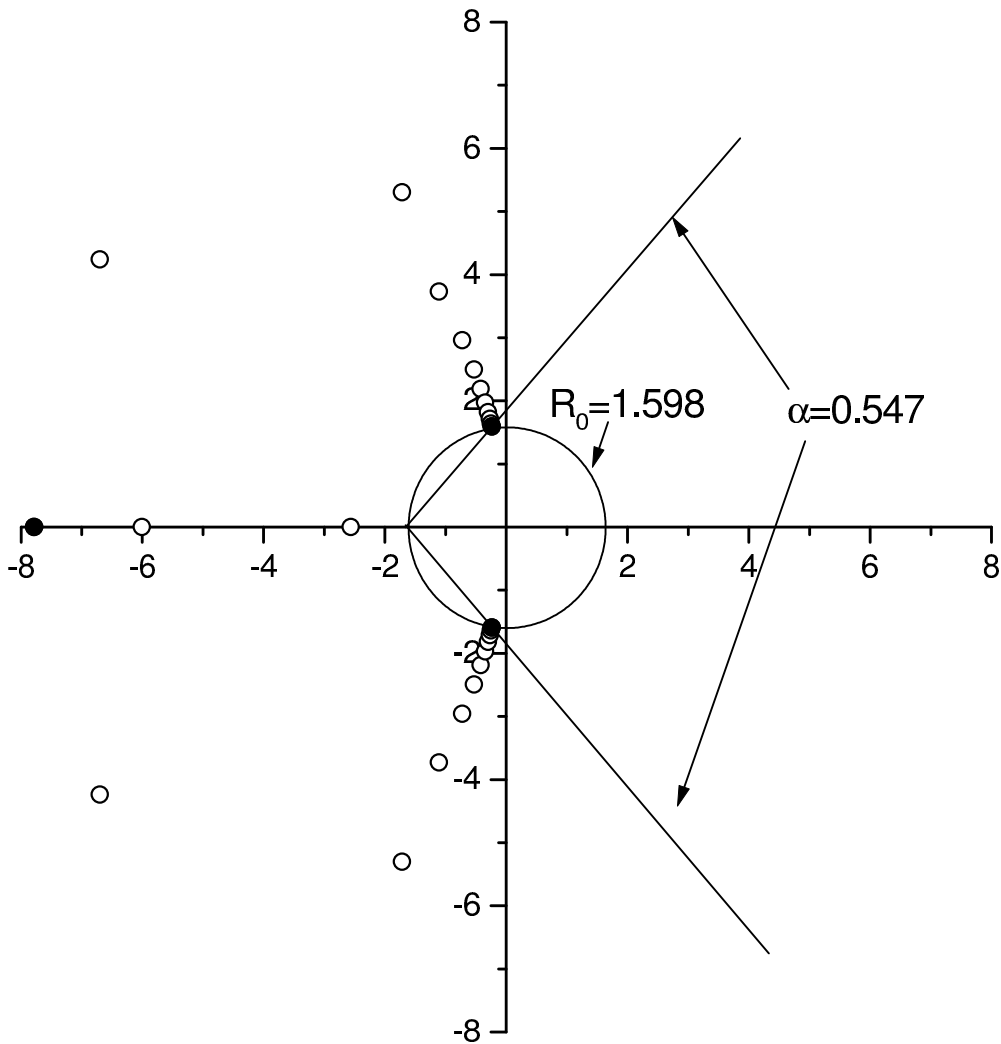}
\end{center}
\caption{ Singularities (small
circles) in the complex plane of $z$ of the solution of the ODE (%
\protect\ref{eq00bis}) for $k\approx -0.52541$ and $M=51$. The three solid
small circles indicate the position of the most important stable
singularities of the Pad\'{e} approximants. The two open circles on the real
axis are spurious (unstable) singularities. The fact that the stable (black
circle) real singularity is far from the origin allows a freedom in the
choice of the parameters $R$ and $\protect\alpha $ of the conformal mapping
method and in particular $R$ may take on a value larger than the radius $%
R_{0}$ of convergence of the Maclaurin series, see text for more details.}
\label{fig9}
\end{figure}

This is sufficient to determine values for the two parameters of the
conformal mapping method:%
\begin{equation}
R=1.59810,\quad \alpha =0.547\,.  \label{eq:AhmadMaxRal}
\end{equation}

However, because the first real singularity is far from the origin (see
figure \ref{fig9}), the largest possible value of the parameter $R$ of the
mapping method does not coincide with the radius $R_{0}=1.59810$ of
convergence of the Maclaurin series. This is true each time the closest
singularity is not located on the negative real axis (the maximal value of $%
R $ is always associated with the closest negative real singularity). But
the largest possible value of $R$ is not necessarily its optimal value as
parameter in the conformal mapping method because it induces a decrease of $%
\alpha $ compare to the choice $R=R_{0}.$ This is why we have kept the
values (\ref{eq:AhmadMaxRal}) with which we have obtained the following
estimate of $k^{\ast }$ (for $M=80$):%
\begin{equation*}
k^{\ast }=-0.52541018\pm 1\times 10^{-8}\,. 
\end{equation*}%
We have observed that the ``explicit'' mapping procedure has appeared more
efficient than the ``minimal'' one which has given a slightly less accurate
estimate of $k^{\ast }$. This effect may be associated with the remark done
above concerning the greater difficulty of detecting an essential
singularity located at infinity compared to a pole or a cut. We note also
that despite a less favorable distribution of singularities the mapping
method has, this time, clearly given a better estimate than the estimate (%
\ref{eq:Ahmadkstarpade}) obtained with the Pad\'{e}-Hankel method
(confirming again the nature of this method, see section \ref{PHMovable}).

Actually, there is a much more efficient method to solve our problem. It is
based on the following change of variable as proposed in \cite{6634}:%
\begin{equation*}
w=e^{-z}\,, 
\end{equation*}%
which transforms the original problem (\ref{eq00bis}--\ref{eq02bis}) into:%
\begin{eqnarray*}
w^{2}\,\bar{g}^{\prime \prime }(w)\,\left[ 1+w^{2}\,\bar{g}^{\prime }(w)^{2}%
\right] +w\,\bar{g}^{\prime }(w)\,\left[ 1+w^{2}\,\bar{g}^{\prime }(w)^{2}%
\right] && \\
-\bar{g}(w)\,\left[ 1+\frac{1}{3}w^{2}\,\bar{g}^{\prime }(w)^{2}\right]
&=&0\,, \\
\bar{g}\left( 0\right) &=&0\,, \\
\bar{g}\left( 1\right) &=&\sqrt{b_{1}c}\,.
\end{eqnarray*}%
with a new connection parameter:%
\begin{equation*}
\bar{g}^{\prime }\left( 0\right) =G\,. 
\end{equation*}

On expressing the generic solution $\bar{g}\left( w\right) =g\left[ z\left(
w\right) \right] $ as a Maclaurin series about $w=0$, the first terms of
this expansion read [see (\ref{eqserinfini0})]:%
\begin{equation}
\bar{g}\left( w\right) =G\,w-\frac{1}{12}G^{3}\,w^{3}+\frac{19}{432}%
G^{5}\,w^{5}-\frac{577}{15552}\,G^{7}w^{7}+\frac{4057}{103680}%
\,G^{9}w^{9}+O(\,w^{11})\,.  \label{eq:gbarw}
\end{equation}

It is a matter of fact that the (simple) sum of the series (\ref{eq:gbarw})
converges in the range $w\in \left[ 0,1\right] $. Then by imposing that this
simple sum satisfies the required condition at the second boundary ($w=1$%
), namely:

\begin{equation*}
\bar{g}^{\ast }\left( 1\right) =g^{\ast }\left[ 0\right] =\sqrt{b_{1}c}=%
\sqrt{\frac{3}{10}}\,, 
\end{equation*}%
we have very easily obtained, with 60 terms in the series, the following
estimate:%
\begin{equation}
G^{\ast }=0.56044886606934678\pm 1\times 10^{-17}\,.  \label{eq:GstarAhmad}
\end{equation}

Using the relation:%
\begin{equation*}
g^{\prime }\left( z\right) =-w\,\bar{g}^{\prime }\left( w\right) \,, 
\end{equation*}%
it comes:%
\begin{equation*}
k=-\,\bar{g}^{\prime }\left( 1\right) =-G+\frac{1}{4}G^{3}-\frac{95}{432}%
G^{5}+\cdots \,, 
\end{equation*}%
which, for $G=G^{\ast }$ as expressed in (\ref{eq:GstarAhmad}), finally
gives the following estimate for the quantity originally of interest:%
\begin{equation*}
k^{\ast }\approx -0.525410175091336\,. 
\end{equation*}

Using Pad\'{e} approximants on the Maclaurin series in powers of $w$ (\ref%
{eq:gbarw}), we have verified that, for $G=G^{\ast }$ as given by (\ref%
{eq:GstarAhmad}), the singularities\ of $\bar{g}^{\ast }\left( w\right) $
are disposed in the complex plane of the variable $w$ in such a manner that
the radius of convergence of the series is actually larger than one.
Consequently, for the value of $A$ considered [see eqs. (\ref{eq01bis}, \ref%
{eqA})], the series actually converges and the recourse to an analytic
continuation, such as a Pad\'{e} approximant, as done in \cite{6634}, is not
necessary to obtain an explicit analytical form of the solution. However,
when $A$ grows, this radius of convergence decreases and becomes smaller
than one\ for some $A<1$ so that the simple sum no longer converges at $w=1$%
. It is very likely that for those values of $A$, the use of one of the two
methods considered in this paper and applied to the series in powers of $w$,
would have some efficiency.

\subsection{The Falkner-Skan flow equation \label{FS}}

In this section we consider the boundary layer Falkner-Skan equation for
wedge. By choosing an especial case of magnetic field, the boundary layer
similarity equation obtained is (Abbasbandy and Hayat \cite{6440}):%
\begin{eqnarray}
g^{\prime \prime \prime }+gg^{\prime \prime }+\beta \left( 1-g^{\prime
2}\right) -\lambda ^{2}\left( g^{\prime }-1\right) &=&0\,,  \label{eq:0} \\
g\left( 0\right) =0,\quad g^{\prime }\left( 0\right) &=&0\,,
\label{eq:CondMHD2} \\
g^{\prime }\left( \infty \right) &=&1\,.  \label{eq:CondMHD3}
\end{eqnarray}

The connection parameter is defined as:%
\begin{equation*}
k=g^{\prime \prime }\left( 0\right) \,. 
\end{equation*}

The Falkner-Skan ODE (\ref{eq:0}) is much more complicated than the previous
examples. Its complete discussion is out of the scope of this article and,
for the sake of our illustration, we limit ourselves to the case $\beta =4/3$
and $\lambda =2$ already studied in \cite{6440} and for which the first
terms of the Maclaurin series of $g\left( z\right) $ are:

\begin{equation*}
g\left( z\right) =\frac{k}{2}-\frac{8}{9}\,z+\frac{k}{6}\,z^{2}+\frac{1}{360}%
\left( 5k^{2}-64\right) \,z^{3}-\frac{k}{135}\,z^{4}+\frac{4\left(
3k^{2}+4\right) }{2835}\,z^{5}+O\left( z^{6}\right) \,. 
\end{equation*}

One may verify that (\ref{eq:0}) admits the following three-parameter
movable singularity as local solution for $z$ close to $z_{0}$  (see appendix A):%
\begin{eqnarray}
g_{\mathrm{sing}}\left( z\right) &\simeq &-\frac{6}{\beta -2}\frac{1}{z-z_{0}%
}+\bar{C}_{1}\left( z-z_{0}\right) ^{\alpha _{+}}+\bar{C}_{2}\left(
z-z_{0}\right) ^{\alpha _{-}}\,,  \label{eq:FSsing1} \\
\alpha _{\pm } &=&\frac{5\beta -4\pm \sqrt{25\beta ^{2}-16\beta -32}}{%
2(\beta -2)}\,,  \notag
\end{eqnarray}%
in which $z_{0}$, $\bar{C}_{1}$ and $\bar{C}_{2}$ are arbitrary constants.

For the expansion (\ref{eq:FSsing1}) to make sense, one must have $\mathrm{Re%
}\left( \alpha _{\pm }\right) >-1$, which occurs for:

\begin{equation*}
\beta \leq \frac{4}{25}\left( 2-3\sqrt{6}\right) \approx -0.855755\quad 
\mathrm{or\quad }\beta >2\,. 
\end{equation*}

These constraints exclude the value of interest $\beta =4/3$, for which we
have:%
\begin{equation*}
\alpha _{\pm }=-2\pm i\sqrt{5}\,, 
\end{equation*}%
and, in the circumstances, the expansion (\ref{eq:FSsing1}) does not locally
represent the general solution.

There is, however, another movable singular solution which is of interest to
us. It involves only two arbitrary parameters (thus it is not a general
solution), and locally about $z_{0}$ behaves as\footnote{%
Since this is a local expression of a particular solution, it may well have
only poles (no logarithm) without implying that the Painlev\'{e} property is
satisfied.}:

\begin{eqnarray}
g_{\mathrm{sing}}\left( z\right) &\simeq &\frac{\text{$C_{1}$}}{\left(
z-z_{0}\right) ^{3}}+\frac{10}{z-z_{0}}-\left( z-z_{0}\right) \left( \frac{1%
}{3\text{$C_{1}$}}+\frac{3\lambda ^{2}}{20}\right)  \notag \\
&&+\frac{\left( z-z_{0}\right) ^{3}\left( 140-27\text{$C_{1}$}\lambda
^{2}\right) }{378\text{$C_{1}$}^{2}}+O\left( \left( z-z_{0}\right)
^{5}\right) \,.  \label{eq:FSsing2}
\end{eqnarray}

We observe that this expression is singular when $C_{1}\rightarrow 0$ and as
in section \ref{ThirdGrade} [eq. (\ref{eq:AhmadSing})] this presumably
indicates the existence of a unique solution defined in $\mathcal{D}^{+}$
(the envelope of the singular solutions) which could be $g^{\ast }\left(
z\right) $.

Eq.(\ref{eq:0}) admits also two particular exact solutions of no interest to
us since they do not correspond to the required boundary conditions:%
\begin{eqnarray*}
g_{1}\left( z\right) &=&z+B_{1}, \\
g_{2}\left( z\right) &=&-\frac{\beta +\lambda ^{2}}{\beta }z+B_{2}\,.
\end{eqnarray*}

These two linear behaviors appear also as leading terms of two asymptotic
expansions ($z\rightarrow \infty $). In fact, for the value of interest%
\footnote{%
For $\beta <0$, there is an additional arbitrary correction term to (\ref%
{eq:FSasy}) of the form $\bar{G}\,z^{2\beta +1}$.} $\beta =4/3$, (\ref{eq:0}%
) admits, asymptotically when $z\rightarrow \pm \infty $, a one-parameter
solution which is of interest to us:

\begin{eqnarray}
&&g\left( z\right) \underset{z\rightarrow \pm \infty }{\simeq }%
z+Gz^{-\lambda ^{2}-2\beta -2}e^{-\frac{z^{2}}{2}}\left[ 1-\frac{\left(
\lambda ^{2}+2\beta +2\right) \left( \lambda ^{2}+2\beta +3\right) }{2z^{2}}%
+O\left( z^{-4}\right) \right]  \notag \\
&&-\frac{1}{4}G^{2}(\beta -1)z^{-2\lambda ^{2}-4\beta -5}e^{-z^{2}}\left[
1+O\left( z^{-2}\right) \right] +O(z^{a}e^{-3\frac{z^{2}}{2}})\,,
\label{eq:FSasy}
\end{eqnarray}%
in which $a$ has to be defined in terms of $\beta $ and $\lambda $. It is
potentially the asymptotic behavior of the solution $g^{\ast }\left(
z\right) $ for [see eq.\ (\ref{eq:CondMHD3})].

For some values of $\beta $ and $\lambda $ (not for the value $\beta =4/3$
however) it exists another particular asymptotic solution (corresponding to
the second exact linear solution) that behaves as:%
\begin{equation*}
g\left( z\right) \underset{z\rightarrow \pm \infty }{\simeq }-\frac{\beta
+\lambda ^{2}}{\beta }\,z\,, 
\end{equation*}%
with exponential corrections. This behavior does not satisfy the condition (%
\ref{eq:CondMHD3}) at the second boundary and does not exist for\ $\beta
=4/3 $. We shall not discuss it further.

The existence of the one-parameter asymptotic solution satisfying the
required condition (\ref{eq:CondMHD3}) at the second boundary indicates that 
$g^{\ast }\left( z\right) $, if it exists, is likely to be unique. It is
probably the envelope of the family of the two-parameter solutions having
locally the movable singularity (\ref{eq:FSsing2}).

It is thus not surprising that the ``minimal'' Pad\'{e}-Hankel method
succeeds in determining a unique value of $k^{\ast }$ \cite{6440} (with $%
M=30 $). We have redone the calculation for a greater $M=51$ and obtain the
following improved estimate (for other values of the parameters $\beta $ and 
$\lambda $, see \cite{6440}):%
\begin{equation}
k^{\ast }=2.43949894\pm 1\times 10^{-8}\,.  \label{eq:FSPH}
\end{equation}

\begin{figure}[tbp]
\begin{center} 
\includegraphics*{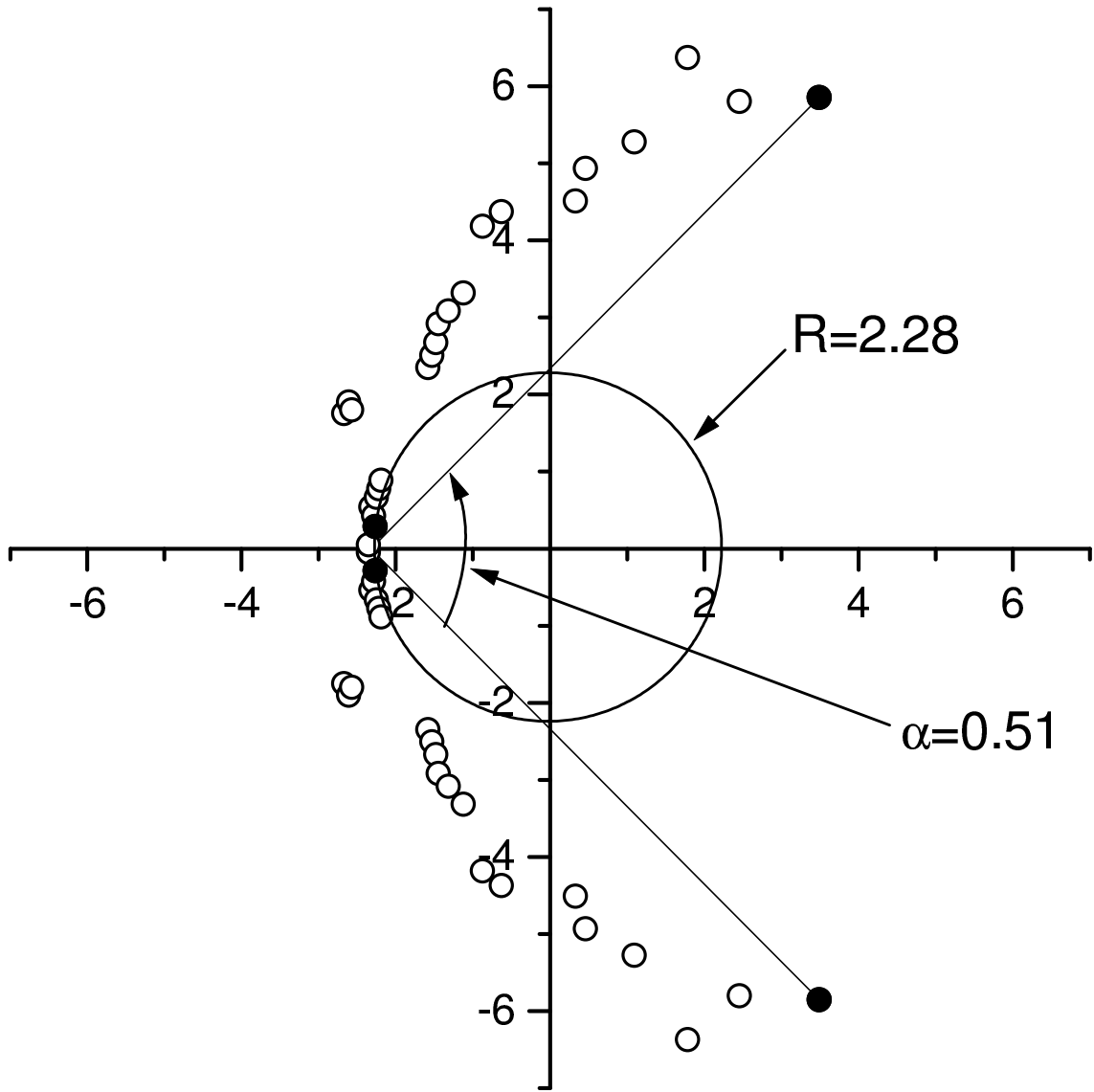}
\end{center}
\caption{Singularities (small
circles) in the complex plane of $z$ of the solution of the Falkner-Skan ODE
(\protect\ref{eq:0}) with $\protect\lambda =2$ and $\protect\beta =4/3$, for 
$k\approx 2.43949$ and $M=100$. The four solid small circles indicate the
position of the most important stable singularities of the Pad\'{e}
approximants. The values $R=2.28$ and $\protect\alpha =0.51$ are the
theoretically optimal values of the parameter of the conformal mapping eq. (%
\protect\ref{eq:Map}), see text for more details.}
\label{fig6}
\end{figure}

Using again Pad\'{e} approximants we may have a look at the distribution of
the singularities of $g^{\ast }\left( z\right) $ in the complex plane of $z,$
see figure \ref{fig6}. From this distribution we obtain:%
\begin{equation*}
R\approx 2.28,\quad \alpha \approx 0.51\,. 
\end{equation*}

\begin{figure}[tbp]
\begin{center} 
\includegraphics*{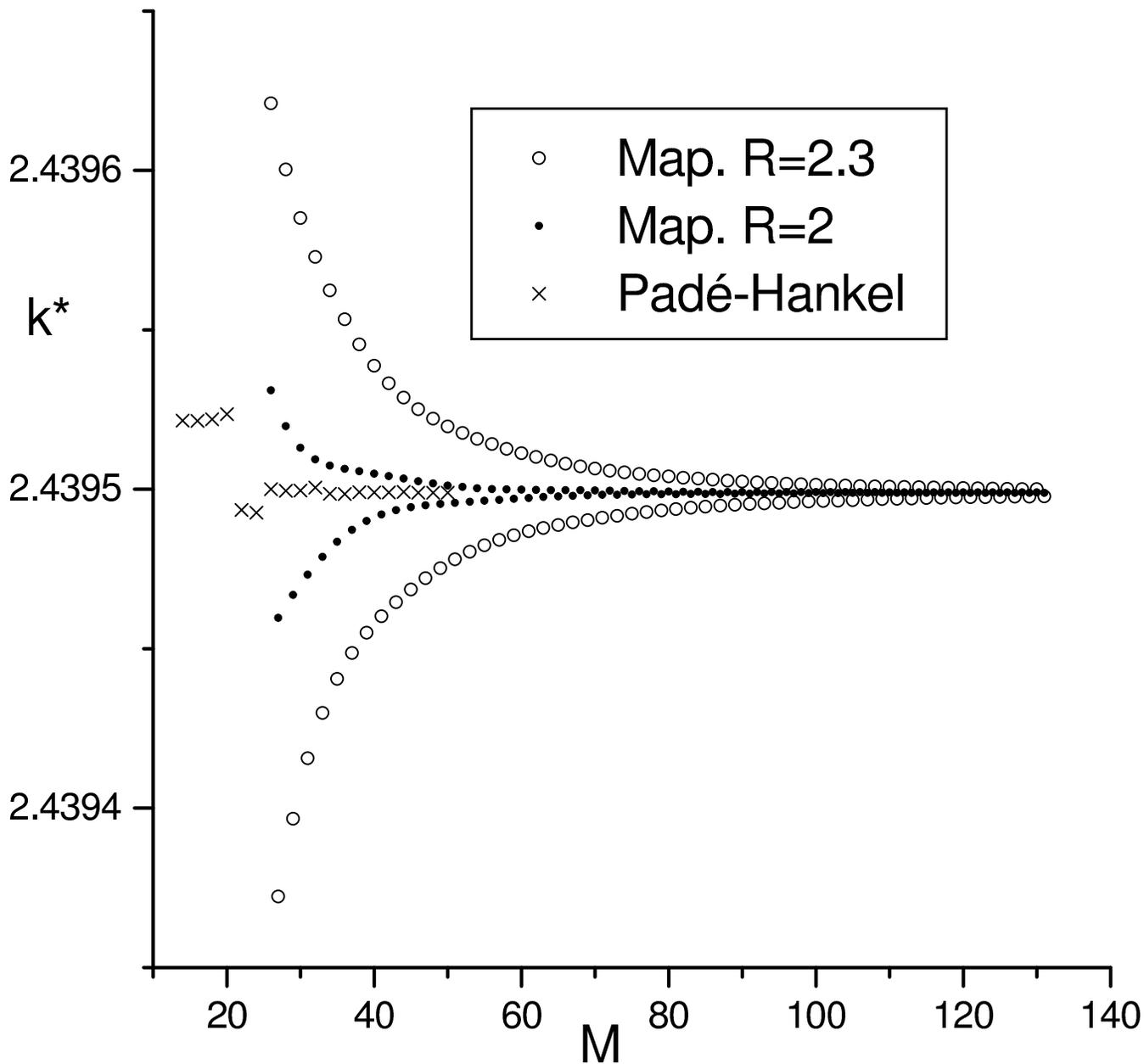}
\end{center}
\caption{[Falkner-Skan ODE (%
\protect\ref{eq:0}) with $\protect\lambda =2$ and $\protect\beta =4/3]$
Comparison of the evolutions with $M$ of the estimate of $k^{\ast }$
according to the method used. Two configurations of the conformal mapping
for a fixed $\protect\alpha =1/2$ (resp. $R=2.3$ and $R=2$) are shown (see
text for more details).}
\label{fig7}
\end{figure}

Using values of $R$ and $\alpha $ close to these estimates, the conformal
mapping method yields convergent procedures. However, even with $M=130,$ the
result does not actually improve the estimate (\ref{eq:FSPH}) obtained by
the Pad\'{e}-Hankel method with $M=51$ (see fig. \ref{fig7}). The better
efficiency of this latter method compared to the primer one is reestablished
and reinforced due to the hardening of the analyticity properties of the
solution.

\subsection{The Blasius Equation\label{Blasius}}

\subsubsection{\protect\bigskip Generalities\label{BlasiusGen}}

The Blasius equation is the Falkner-Skan equation (\ref{eq:0}) with $\beta
=\lambda =0$  \cite{6552}:%
\begin{equation}
g^{\prime \prime \prime }+gg^{\prime \prime }=0\,.  \label{eq:Blasius}
\end{equation}

Let us first \ consider the general BVP of finding a solution to (%
\ref{eq:Blasius}) which satisfies the following conditions:%
\begin{eqnarray}
g\left( 0\right) &=&a,\quad g^{\prime }\left( 0\right) =b\,,
\label{eq:BlasiusCond01} \\
g^{\prime }\left( +\infty \right) &=&B\,,  \label{eq:BlasiusCond03}
\end{eqnarray}%
and the connection parameter $k$ is defined by:%
\begin{equation}
k=g^{\prime \prime }\left( 0\right) \,.  \label{eq:BlasiusCP}
\end{equation}

Eq. (\ref{eq:Blasius}) may be integrated to give:%
\begin{equation}
g^{\prime \prime }\left( z\right) =k\exp \left( -\int_{0}^{z}g\left(
t\right) \right) \,,  \label{eq:sol}
\end{equation}%
consequently $g\left( z\right) $ is either convex if $k>0$, or concave if $%
k<0$ or linear if $k=0$.

For the sake of our illustration, we limit ourselves to the case $B=1$ with $%
b\in \lbrack 0,1)$ which corresponds to convex solutions (for a discussion
of the concave solutions, see for example \cite{6558}).

There are two explicit exact particular solutions to (\ref{eq:Blasius}), the
first one corresponds to the linear case:%
\begin{equation*}
g_{\mathrm{lin}}\left( z\right) =a+b\,z\,, 
\end{equation*}%
and the second one has a movable singularity: 
\begin{equation}
g_{\mathrm{sing}}=\frac{3}{z-z_{0}}\,,  \label{eq:Blasiussing}
\end{equation}%
corresponding to the particular conditions at the origin:%
\begin{equation*}
a=-\frac{3}{z_{0}},\quad b=-\frac{3}{z_{0}^{2}},\quad k=-\frac{6}{z_{0}^{3}}%
\,. 
\end{equation*}

Because the ODE is autonomous, these two kinds of particular solutions
(linear and singular) appear again, either as asymptotic limits when $%
z\rightarrow +\infty $, or as singular limits in the two main kinds of
solution (concave and convex):

\begin{itemize}
\item Similarly to the Falkner-Skan case, there is a three-parameter movable
singularity representing locally the general solution [see eq.(\ref%
{eq:FSsing1}) for $\beta =0$]%
\begin{eqnarray}
g_{\mathrm{sing}}\left( z\right) &\simeq &3\frac{1}{z-z_{0}}+C_{1}\left(
z-z_{0}\right) ^{\alpha _{+}}+C_{2}\left( z-z_{0}\right) ^{\alpha _{-}}\,, \\
\alpha _{\pm } &=&1\pm i\sqrt{2}\,,  \notag
\end{eqnarray}%
which occurs only in the concave case.

\item For large $z$, two asymptotic solutions exist:

\begin{enumerate}
\item a one-parameter asymptotic solution ($\bar{G}_{1}$ is arbitrary):%
\begin{equation*}
g_{\mathrm{asy}}\left( z\right) \underset{z\rightarrow \infty }{\simeq }%
\frac{3}{z}+\frac{\bar{G}_{1}}{z^{2}}+\frac{\bar{G}_{1}^{2}}{3z^{3}}+O\left(
z^{-4}\right) \,, 
\end{equation*}%
which does not correspond to the convex case.

\item a two-parameter asymptotic solution ($G_{1}$ and $G_{2}$ are
arbitrary):
\end{enumerate}
\end{itemize}

\begin{equation}
g_{\mathrm{asy}}\left( z\right) \underset{z\rightarrow \infty }{\simeq }%
G_{1}z+\frac{G_{2}}{z^{2}}e^{-\frac{G_{1}z^{2}}{2}}\,,  \label{eq:Blasiusasy}
\end{equation}%
which is compatible with the convex case if $G_{1}>0$ and candidate to be
the solution of interest if $G_{1}=B=1.$

Notice that (\ref{eq:Blasiusasy}) replaces the one-parameter asymptotic
solution (\ref{eq:FSasy}) of the Falkner-Skan case. This is in agreement
with the disappearance of the particular two-parameter local solution with
movable singularity (\ref{eq:FSsing2}) and, consequently, of their envelope
and the possibility of having an isolated (unique) solution defined in $%
\mathcal{D}^{+}$ (at least in the convex case). Let us confirm this by the
following considerations.

In the convex case that we consider, we have $g\left( t\right) \geq a+bt$
and \ from (\ref{eq:sol}) we can write:%
\begin{equation*}
0<g^{\prime \prime }\left( z\right) \leq k\,e^{-a\,z-\frac{1}{2}b\,z^{2}}\,, 
\end{equation*}%
consequently the domain of definition of these solutions is $\mathcal{D}^{+}$
and there is a continuum of such solutions corresponding to any $k>0$. Of
course only one of them gives $g^{\prime }\left( +\infty \right) =1$, but if
that latter condition is not explicitly imposed, it is unlikely that the
right value of $k^{\ast }$ be determined. Consequently, the ``minimal''
procedures are inefficient here. This is the reason why the Pad\'{e}-Hankel
method has failed to solve the Blasius problem \cite{6190}.

\subsubsection{\protect\bigskip The Blasius problem}

The Blasius problem consists in finding the solution of (\ref{eq:Blasius})
which satisfies the following boundary conditions ($B=1$, $a=b=0$) [let us mention the interesting Boyd's review on this problem \cite{7095}]:%
\begin{eqnarray}
g\left( 0\right) &=&0,\quad g^{\prime }\left( 0\right) =0\,,
\label{eq:BlasiusCond1} \\
g^{\prime }\left( \infty \right) &=&1\,.  \label{eq:BlasiusCond3}
\end{eqnarray}

With the conditions (\ref{eq:BlasiusCond1}) at the origin, the function $%
g\left( z\right) $ appears to be of the form:%
\begin{equation}
g\left( z\right) =z^{2}u\left( z^{3}\right) \,.  \label{eq:BlasiusChange1}
\end{equation}

Indeed Blasius \cite{6552} has given the general expression of the Maclaurin
series of $g\left( z\right) $ corresponding to the initial conditions (\ref%
{eq:BlasiusCond1}), it reads:%
\begin{eqnarray}
g\left( z\right) &=&\sum_{i=0}^{\infty }\left( -1\right) ^{i}\frac{%
A_{i}k^{i+1}}{\left( 3i+2\right) !}z^{3i+2}\,,  \label{eq:BlasiusSer1} \\
A_{i} &=&\left\{ 
\begin{array}{c}
1\qquad i=0\cup i=1 \\ 
\sum_{j=0}^{i-1}\frac{\left( 3i-1\right) !}{\left( 3j\right) !\left(
3i-3j-1\right) !}A_{j}A_{i-j-1}\qquad i\geq 2%
\end{array}%
\right. \,,  \label{eq:BlasiusSer2}
\end{eqnarray}%
which corresponds to the following first explicit terms:

\begin{equation*}
g\left( z\right) =\frac{k}{2}z^{2}-\frac{k^{2}}{120}z^{5}+\frac{11k^{3}}{%
40320}z^{8}-\frac{5k^{4}}{532224}z^{11}+\frac{9299k^{5}}{29059430400}%
z^{14}+O\left( z^{17}\right) \,. 
\end{equation*}

Moreover the particular form of (\ref{eq:Blasius})] induces the following
scale invariance \cite{6572}:%
\begin{eqnarray}
z &\rightarrow &\bar{z}=\frac{z}{\sigma }\,,  \label{eq:zbar} \\
g &\rightarrow &\,\bar{g}=\sigma \,g\left( \sigma \bar{z}\right) \,,
\label{eq:gbar}
\end{eqnarray}%
so that, without changing the conditions (\ref{eq:BlasiusCond1}), the new
connection parameter $\bar{k}=\bar{g}^{\prime \prime }\left( 0\right)
=\sigma ^{3}k$ may be set a priori equal to any value, say $q_{0}$, in the
Maclaurin series of the Blasius problem, i.e.: 
\begin{equation*}
\bar{k}=q_{0}\Longrightarrow \sigma =\left( \frac{q_{0}}{k}\right) ^{1/3}\,, 
\end{equation*}%
in which case, the value of $\bar{g}^{\prime }\left( \infty \right) $ at
infinity reads:%
\begin{equation*}
\bar{g}^{\prime }\left( \infty \right) =\left( \frac{q_{0}}{k}\right)
^{2/3}g^{\prime }\left( \infty \right) \,. 
\end{equation*}%
Hence, for the required condition (\ref{eq:BlasiusCond3}) at infinity, we
simply read the value of $k^{\ast }$ as the result of a simple Cauchy
problem with the conditions (\ref{eq:BlasiusCond1}) and $g^{\prime \prime
}\left( 0\right) =q_{0}$:%
\begin{equation}
k^{\ast }=q_{0}\,\left[ \bar{g}^{\prime }\left( \infty \right) \right]
^{-3/2}\,.  \label{eq:Blasiuskstarq0}
\end{equation}%
We have thus integrated numerically (\ref{eq:Blasius}) for $\bar{k}=1$
(using Mathematica) and obtained (see also Boyd \cite{7089})\footnote{%
There are several normalizations used in the literature. The most common is
to choose a function $f\left( z\right) =2g\left( z\right) $ with $B=1$ and
the connection parameter $k_{0}=f^{\prime \prime }\left( 0\right) $, in
which case one has $k_{0}^{\ast }=0.33205733621519630$ \cite{7089}. For the
choice $g\left( z\right) $ we have made, we get $k^{\ast }=$ $k_{0}^{\ast }$
if one fixes $B=2$ (the genuine Blasius problem). With the present choice of 
$B=1$, using the scale invariant properties (\ref{eq:gbar}), we have $%
k^{\ast }=\sqrt{2}k_{0}^{\ast }$ which corresponds to (\ref{eq:BlasiusKstar}%
). As for the radius of convergence $R_{0}=5.6900380545$ of the Maclaurin
series in powers of $z$ given in \cite{7089}, it is related to our $R=65.113$
for the series in powers of $x=z^{3}$ via the relation $R=\left( R_{0}/\sqrt{%
2}\right) ^{3}=65.13291560$.\label{footblas}}:%
\begin{equation}
k^{\ast }\approx 0.469599988361013304509\,.  \label{eq:BlasiusKstar}
\end{equation}

We may also sum the Maclaurin series using a Pad\'{e} approximant or an
adequate conformal mapping. To this end, it is useful to perform the change
of function $g\rightarrow u$ of (\ref{eq:BlasiusChange1}) and to work with
the new independent variable 
\begin{equation*}
x=z^{3}\,. 
\end{equation*}

If we note $\bar{u}\left( x\right) $ the function corresponding to $\bar{g}%
\left( z\right) $ introduced above with $\bar{g}^{\prime \prime }\left(
0\right) =\bar{k}=1$, then we have:%
\begin{equation}
\left[ \bar{g}^{\prime }\left( z\right) \right] ^{3}=x\left[ 2\bar{u}\left(
x\right) +3x\bar{u}^{\prime }\left( x\right) \right] ^{3}\,.
\label{eq:Blasiusgbarprime}
\end{equation}

One may then perform a diagonal Pad\'{e} approximant of the corresponding
truncated series and obtain an estimate of $k^{\ast }$ by considering the
following limit of the rational fraction:%
\begin{equation}
k^{\ast }=\left\{ \lim_{x\rightarrow \infty }\mathrm{Pad\acute{e}}\left( x%
\left[ 2\bar{u}\left( x\right) +3x\bar{u}^{\prime }\left( x\right) \right]
^{3}\right) \right\} ^{-1/2}\,.  \label{eq:BlasiusPAD}
\end{equation}%
This way we have obtained (with $M=200$):%
\begin{equation}
k^{\ast }\approx 0.46959998836100\,,  \label{eq:BlasiusKstarPade}
\end{equation}%
but the observed convergence is not limpid.

We may also extract, from the Pad\'{e} approximant of $u\left( x\right) $
with a given $k$, the corresponding distribution of singularities in the
complex plane of $x$. For the value $k=0.4696$ close to (\ref%
{eq:BlasiusKstarPade}), we obtain a distribution of singularities located in
the lhs half plane with a cut on the real axis starting at $x=-65.133$ and
some singularities elsewhere which induces the following values for the
parameters of the conformal mapping:%
\begin{equation*}
R=65.133,\quad \alpha =1.57\,. 
\end{equation*}

As noted in footnote \ref{footblas}, this value of $R$ agrees with the
preceding estimate of \cite{7089} in which reference one may find a
discussion of the distribution of the singularities in a particular choice
of normalization for which $\alpha =3/2$. For other normalizations and other
values of $k$, the estimates for $\alpha $ and $R$ change. We have performed
a conformal mapping on a series with $M=590$ for a fixed $k=1$ and obtained
after a single sum of the series [using (\ref{eq:Blasiuskstarq0}) with $%
q_{0}=1$]:%
\begin{equation*}
k^{\ast }\approx 0.46959998836102\,. 
\end{equation*}

That estimate is not more accurate than (\ref{eq:BlasiusKstarPade}), showing
again the great efficiency of the Pad\'{e} approximants (though it would be
difficult to improve the result whereas this would be relatively easy with
conformal mappings).

For illustrative purposes we present Table \ref{Tab1} in which we present
our results compared to various estimates of $k^{\ast }$ encountered in the
literature.

\begin{center}
\begin{table}[tbp] \centering%
\begin{tabular}{lcl}
\hline\hline
Source & $k^{\ast }$ & Meth \\ \hline
present work & \multicolumn{1}{l}{0.469599988361013304509} & Integ \\ 
Boyd \cite{7089} & \multicolumn{1}{l}{0.46959998836101328} & Integ \\ 
present work & \multicolumn{1}{l}{0.46959998836102} & MAP (590) \\ 
present work & \multicolumn{1}{l}{0.46959998836100} & PAD (200) \\ 
Fazio \cite{6569} & \multicolumn{1}{l}{0.469599988361} & SHO \\ 
Asaithambi \cite{6557} & \multicolumn{1}{l}{0.46959998847} & SHO \\ 
Parlange \cite{6570} & \multicolumn{1}{l}{0.46959999} & Integ \\ 
Yu-Chen \cite{6554} & \multicolumn{1}{l}{0.4695997} & DTM \\ 
Liao \cite{6559} & \multicolumn{1}{l}{0.469500} & HAM \\ 
Howarth \cite{6571} & \multicolumn{1}{l}{0.46960} & SHO \\ 
Blasius \cite{6552} & \multicolumn{1}{l}{0.4695} & Matching%
\end{tabular}%
\caption{(Blasius problem) Estimates of k* of the present study compared to other estimates encountered in the literature. 
The last column on the right indicates the method used. ``Matching'': matching of the sum of the Maclaurin 
series with the sum of an asymptotic expansion; ``SHO'': shooting method (Cauchy+Newton); ``Integ'': a single
integration (Cauchy); ``HAM'': homotopy analysis method; ``DTM'': Differential Transformation Method; 
``PAD (200)'': Padé approximants on the Maclaurin series with $M=200$ using eq.(\ref{eq:BlasiusPAD});
``MAP (590)'': a conformal mapping on the Maclaurin series with $M=590$ using eq.(\ref{eq:Blasiuskstarq0}).
}\label{Tab1}%
\end{table}%
\end{center}

The Blasius problem is finally easy to solve because of the conjunction of
the scaling invariance and of the initial conditions (\ref{eq:BlasiusCond1})
which induce the effective independent variable $x=z^{3}$ and increases by a
factor three the angle of the domain of analyticity of the solution. This
favorable conjonction no longer exists with different initial conditions as
shown in the following subsection.

\subsubsection{A more general convex solution \label{Convex}}

Let us consider the BVP of finding a solution to (\ref{eq:Blasius}%
) which satisfies the following conditions (a similar problem has been studied by Allan in \cite{6625}):%
\begin{eqnarray}
g\left( 0\right) &=&0,\quad g^{\prime }\left( 0\right) =b\,,
\label{eq:Convex-1} \\
g^{\prime }\left( +\infty \right) &=&1\,,  \label{eq:Convex-2}
\end{eqnarray}%
as explained in section \ref{BlasiusGen}, the corresponding solution is
convex provided that $b<1$, in which case the associated connection
parameter $k$ defined in (\ref{eq:BlasiusCP}) takes on a positive (still
unknown) value $k_{0}^{\ast }$.

\paragraph{A simple dichotomy algorithm}

For the sake of our illustration, we first consider the case $b=1/4$.

Before considering any series expansion, let us show an interesting aspect
of the invariance of (\ref{eq:Blasius}) under the rescaling (\ref{eq:gbar}, %
\ref{eq:zbar}). Though the BVP could no longer be transformed into
a simple Cauchy problem, we may manage an easy dichotomy procedure to
determine $k_{0}^{\ast }$.

Denoting by $\left[ \sigma =1:\left\{ b,k_{0}^{\ast };1\right\} \right] $
the original BVP corresponding to the boundary conditions (\ref%
{eq:Convex-1}, \ref{eq:Convex-2}), then by virtue of the scaling invariance,
the following equivalence stands:%
\begin{equation*}
\left[ \sigma =1:\left\{ b,k_{0}^{\ast };1\right\} \right] \equiv \left[
\sigma =\left( k_{0}^{\ast }\right) ^{1/3}:\left\{ b\left( k_{0}^{\ast
}\right) ^{-2/3},1;\left( k_{0}^{\ast }\right) ^{-2/3}\right\} \right] \,. 
\end{equation*}

Consequently, to determine the value $k_{0}^{\ast }$ (associated to $b$), we
may proceed as follows. Choosing an arbitrary number $b_{0}$, a single
integration of the ODE (\ref{eq:Blasius}) with the initial condition $%
g\left( 0\right) =0,\quad g^{\prime }\left( 0\right) =b_{0},\quad g^{\prime
\prime }\left( 0\right) =1$ provides the value $g^{\prime }\left( \infty
\right) =B_{0}$ which should be equal to $b_{0}/b$ otherwise one tries the
new value $g^{\prime }\left( 0\right) =\frac{b_{0}+b\,B_{0}}{2}$ and so on
and so forth until $g^{\prime }\left( \infty \right) \approx g^{\prime
}\left( 0\right) /b$, which finally gives $k_{0}^{\ast }\approx \left[
g^{\prime }\left( \infty \right) \right] ^{-3/2}$. Of course the speed of
the method depends on whether the initial value $b_{0}$ is close or not to $%
b\left( k_{0}^{\ast }\right) ^{-2/3}$. By this dichotomy procedure we have
obtained the following estimate for $b=1/4$:%
\begin{equation*}
k_{0}^{\ast }=0.429540737652735\pm 1\times 10^{-15}\,.
\end{equation*}%

At this stage, we already mention that the Taylor series methods, which we
are presently interested in, have failed in providing any reliable estimate
of $k_{0}^{\ast }$ while a [10,10]-Homotopy Pad\'{e} method (for an
introduction see \cite{7167}) gives, with a relative easiness, the estimate $%
k_{0}^{\ast }=0.429543$. Let us look at the
reasons why the Taylor-series-based methods are not efficient there.

\paragraph{Taylor series and the analyticity properties of a solution \label%
{ConvexAnal}}

We already know that the ``minimal'' procedures (in particular the Pad\'{e}%
-Hankel method) cannot work because of a continuum of solutions defined in $%
\mathcal{D}^{+}$ (see section \ref{BlasiusGen}). In addition the
singularities in the complex plane of the independent variable $z$ of the
solution cannot be moved toward the left-hand side as in the
Blasius problem. Consequently some of them are located in the region $\func{%
Re}\left( z\right) >0$ in such a way to limit the efficiency of the Pad\'{e}
and conformal mapping methods. With a view to an illustration of the origin
and the extent of the difficulties encountered, we give up solving the BVP
and turn our attention to the Cauchy problem associated to the
following initial values:%
\begin{equation*}
g\left( 0\right) =0\,,\quad g^{\prime }\left( 0\right) =1/2\,,\quad
g^{\prime \prime }\left( 0\right) =1\,, 
\end{equation*}%
for which a single integration provides the asymptotic value:%
\begin{equation}
g^{\prime }\left( \infty \right) \approx 1.78283921978496662798847022654\,.
\label{eq:gpinf}
\end{equation}

The first terms of the corresponding Maclaurin series read:

\begin{equation}
g\left( z\right) =\frac{z}{2}+\frac{z^{2}}{2}-\frac{z^{4}}{48}-\frac{z^{5}}{%
120}+\frac{z^{6}}{960}+\frac{11}{10080}\,z^{7}+\frac{73}{322560}\,z^{8}-%
\frac{43}{483840}\,z^{9}+O\left( z^{10}\right) \,.  \label{eq:Convex-series}
\end{equation}

\begin{figure}[tbp]
\begin{center} 
\includegraphics*{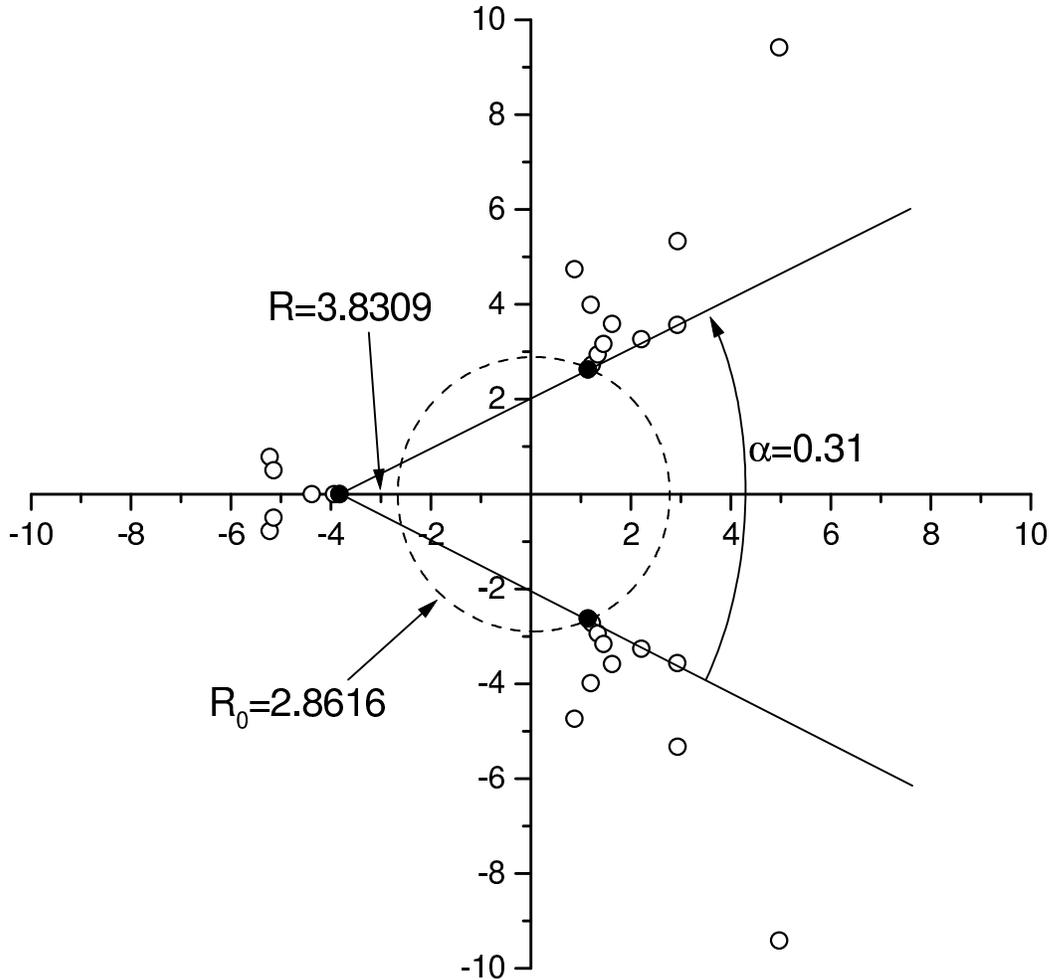}
\end{center}
\caption{Singularities (small
circles) in the complex plane of $z$ of the solution of the Blasius ODE for
the initial conditions (\protect\ref{eq:Convex-1}) with $b=1/2$ and $%
g^{\prime \prime }\left( 0\right) =1$ (from the Pad\'{e} approximant
[35,35], discarding only some spurious singularities). The small solid
circles indicate the positions of the stable singularities the closest to
the origin. The large circle of radius $R_{0}=2.8616$ indicates the domain
of convergence of the Maclaurin series (\protect\ref{eq:Convex-series}) in
powers of $z$. The theoretical optimal values of the parameters of the
mapping method are rather $R=3.8309$ and $\protect\alpha =0.31$. Farther on
the rhs, singularities (not yet completely stable) may perturb the
convergence of the mapped series.}
\label{fig91}
\end{figure}

By performing diagonal Pad\'{e} approximants on this series, we obtain, the
distribution of singularities as shown in fig. \ref{fig91}. From this we
observe that:

\begin{enumerate}
\item the largest possible value of the parameter $R$ of the mapping method
does not coincide with the radius of convergence of the Maclaurin series
(noted $R_{0}$ in fig. \ref{fig91}). But, this time the radius of
convergence $R_{0}$ cannot be chosen as reference to determine the optimal
value of $\alpha $ as it would follow from the idealized scheme of fig. 1 of 
\cite{6319}, this is because:

\item farther singularities in the half plane $\func{Re}\left( z\right) >0$
would prevent the convergence of the mapping method, so that

\item the\ effective maximal optimal value of $\alpha $ ($0.31$) is small
and can be significantly enlarged only if $R$ is considerably diminished.
\end{enumerate}

\begin{figure}[tbp]
\begin{center} 
\includegraphics*{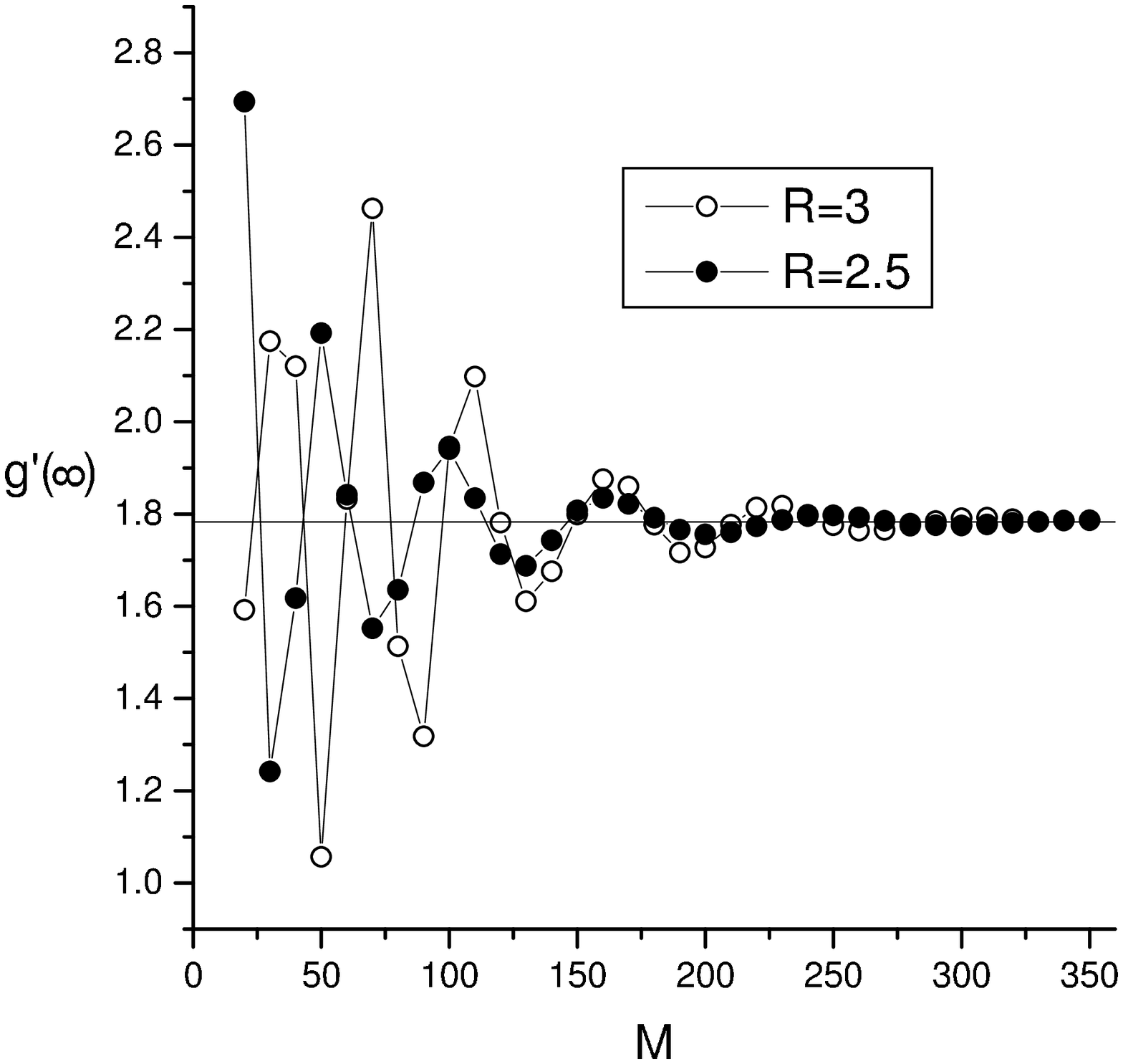}
\end{center}
\caption{[Cauchy problem for
the Blasius ODE (\protect\ref{eq:Blasius}) with initial conditions $a=0$, $%
b=1/2$ and $k=1$] Comparison of the evolutions with $M$ of the estimate of $%
g^{\prime }\left( \infty \right) $ by the mapping method for $\protect\alpha %
=0.3$ and two values of $R$. The solid horizontal line indicates the
expected value (\protect\ref{eq:gpinf}). The convergence is slow but
manifest (see text for more details).}
\label{fig92}
\end{figure}

As consequence of this unfavorable singularity distribution, the mapping
method for the optimal values $R=3.8$ and $\alpha =0.3$, converges only very
slowly (see fig. \ref{fig92}) and, with $M=350$, we have obtained the poor
estimate $g^{\prime }\left( \infty \right) =1.78\pm 0.06.$ Smaller values of 
$R$ yield slightly improved convergences which remain slow however.
Nevertheless the convergence (even slow) of the method indicates that the
distribution of singularities of fig. \ref{fig91} is correct. Despite this
fact, Pad\'{e} approximants do not yield any estimate of $g^{\prime
}\left( \infty \right) $.

\section{Summary and conclusion}

We have presented, compared and discussed the efficiency of three
quasi-analytic methods for solving a BVP: the  Taylor series, the Pad\'{e} and
the conformal mapping methods.
After having reminded that the first method 
almost always requires the recourse to analytical continuation procedures to be efficient, our interest has been focused on the latter two.

 We have emphasized that the Pad\'{e}-Hankel method is indeed  the ``minimal'' version of the standard Pad\'{e}
method and explained why it is not always efficient. In particular we have
shown that its efficiency is closely related with the possibility of pushing
to infinity a movable singularity. The conformal mapping always works 
but (obviously by construction) its efficiency depends on the
analytic properties of the solutions sought.

With a view to present a fairly comprehensive discussion, we have successively
considered six different configurations of ODE and initial conditions which
yield more and more constraining analytic properties. We have shown that Pad%
\'{e} approximants may be used to determine correctly the singularities
distribution which allows an efficient determination of the values of the
two parameters $R$ and $\alpha $ of the conformal mapping method [see eq. (%
\ref{eq:Map})]. In each case of ODE considered we have empirically discussed
the conditions of uniqueness of the solution defined in $\mathcal{D}^{+}=\left[
0,+\infty \right) $, this unique solution appears as the envelope of a
family of solutions having moving singularities, when such a family of
solutions does not exist then an infinite number of solutions are defined in 
$\mathcal{D}^{+}$ and ``minimal'' procedures (like the Pad\'{e}-Hankel
method) do not work.

Contrary to the mapping method, the Pad\'{e}-Hankel method (when it works)
is comparatively more and more efficient as the analytic properties harden
but it is limited by its awkwardness. The mapping method is more
sophisticated and smoother. Provided the analytic properties of the solutions are favorable, it is useful when a high accuracy is required.

\bigskip

\textbf{Acknowledgments }We thank B. Boisseau and H. Giacomini for their
useful remarks and comments relative to this work. We are also indebted to the referees for their useful suggestions.

\appendix
\section{Expansions of solutions about particular points}
Asymptotic expansions such as (\ref{eq:Polasy}, \ref{eq:TFasy}, etc...) and expansions about a movable singularity such as (\ref{eq:Polgsing}, \ref{eq:TFgsing}, etc...) may be determined following a similar procedure. The main difference with the expression of the generic solution as a Taylor series is that the terms are not necessarily integer powers of the independent variables.  For example, it may involve also powers of logarithms. Sometimes the leading term is not a power but an exponential or a logarithm. This prevents us from using a general systematic algorithm to get the terms of such expansions.

As an illustration we present below the example of the asymptotic expansion of the solution of the Polchinski fixed point equation. The obtention of the local expansion about a movable singularity is very similar and will not be made explicit here.
\subsection{Asymptotic expansion of the solution of the Polchinski equation}
Let us consider the ODE (\ref{eq:Pol2}). We want to show that it admits the one-parameter solution (\ref{eq:Polasy}) as $z\rightarrow\infty$.
Although not obliged, it is convenient to perform the following change:
\begin{equation}
g(z)=h(y=1/z)
\end{equation}
then (\ref{eq:Pol2}) reads:
\begin{equation}
4 y^3 h''(y)+2 y^2 h'(y)+4 y h(y) h'(y)+y h'(y)-2 h(y)^2+2 h(y)=0
\end{equation}
Assuming that $y$ is small, we try to find a local solution to this ODE under the following form:
\begin{equation}
h_\mathrm{asy}(y)\simeq\chi\, y^\gamma \label{apphasy}
\end{equation}
where $\chi$ and $\gamma$ are the unknowns.
It is easy to see that (\ref{apphasy}) generates three different powers of $y$ in the ODE: $y^{\gamma}$, $y^{2\gamma}$ and $y^{1+\gamma}$. To get a solution (locally valid) one must first determine the leading power for small $y$ and second impose that its coefficient vanishes. These two conditions should determine the values of $\chi$ and $\gamma$.

Since $y$ is small, the smallest power prevails, then we distinguish three possibilities:
\begin{enumerate}
\item $\gamma<0$, the leading power is $2\gamma$, but the vanishing of the global coefficient imposes $\gamma=1/2$ (whatever $\chi$) what is incompatible with the hypothesis.
\item $\gamma>0$, the power $\gamma$ prevails, but the vanishing of the global coefficient imposes $\gamma=-2$ (whatever $\chi$) what is incompatible with the hypothesis.
\item $\gamma=0$, the two powers $\gamma$ and $2\gamma$ are equal and the vanishing of the global coefficient occurs for $\chi=0$ or $\chi=1$.
\end{enumerate}
The leading term of the asymptotic expansion is thus 1. To get the next term we try the following form:
\begin{equation}
h_\mathrm{asy}(y)=1+\chi\, y^\gamma \label{appasycorr}
\end{equation}
in which the second term is smaller than the first one, i.e. $\gamma>0$, for consistency. Introduced into the ODE, this expansion about $y=0$ gives the same set of powers of $y$ as previously. The leading power is thus $\gamma$. The vanishing of the global coefficient implies that $\gamma=2/5$ whatever $\chi$. That is an acceptable solution.

The arbitrariness of the amplitude provides the first arbitrary constant that is noted $G$ in (\ref{eq:Polasy}).

The local expansion of the solution about $y=0$ may be continued by considering the next correction:
\begin{equation*}
h_\mathrm{asy}(y)=1+G\, y^{2/5}+\chi\, y^\gamma
\end{equation*}
and so on and so forth.

In the present case, $G$ is the only arbitrary constant appearing in the asymptotic expansion. It is important to know the maximal number of arbitrary parameters in the solution (sometimes called ``resonances'' in the expansion about a movable point). However, it may occur that the arbitrary constants appear lately in the expansion [e.g. see (\ref{eq:TFgsing})] which, in addition, may be very complicated. There is a convenient way to quickly know the values of $\gamma$ that are associated to the arbitrary constants in the expansion without having to calculate it explicitly. Once the leading term of the expansion is known, one considers the generic correction as in (\ref{appasycorr}) then it suffices to look at the vanishing of the linear-in-$\chi$ contribution to the ODE. The resulting constraint provides an equation for $\gamma$. In the present example there is a unique acceptable solution to this equation: $\gamma=2/5$, there is thus only one arbitrary constant ($G$).

\end{document}